\documentclass[graybox]{svmult}
\usepackage{pifont}
\usepackage{type1cm}       
\usepackage{makeidx}         
\usepackage{graphicx}                           
\usepackage{multicol}       
\usepackage[bottom]{footmisc}
\usepackage{newtxtext}       
\usepackage{newtxmath}

\DeclareMathOperator{\diag}{diag}
\usepackage[style=super]{glossaries}
\makeglossaries
 \usepackage{bbold}
\DeclareMathOperator{\Tr}{Tr}
\usepackage{float}

\makeindex            
              
\makeatletter
\renewenvironment{svgraybox}%
  {\begin{shaded}%
   \list{}{\leftmargin=\z@\rightmargin=\z@\topsep=\z@}%
   \expandafter\item\parindent=\svparindent
   \hskip-\listparindent}
  {\endlist\end{shaded}}
\makeatother
\usepackage[none]{hyphenat}

\begin{document}
\title*{Automatic Target Detection for Sparse Hyperspectral Images}
\author{Ahmad W. Bitar, Jean-Philippe Ovarlez, Loong-Fah Cheong and Ali Chehab}
\institute{Ahmad W. Bitar \at SONDRA Lab/CentraleSup\'elec | Universit\'e Paris-Saclay, 91192 Gif-sur-Yvette, France;
\\
Department of Electrical and Computer Engineering, American University of Beirut, Beirut, Lebanon (\email{ahmad.bitar@centralesupelec.fr; ab76@aub.edu.lb}).
\and Jean-Philippe Ovarlez \at The French Aerospace Lab (ONERA) | Universit\'e Paris-Saclay, 91120 Palaiseau, France;
\\
 SONDRA Lab/CentraleSup\'elec | Universit\'e Paris-Saclay, 91192 Gif-sur-Yvette, France (\email{jean-philippe.ovarlez@onera.fr; jeanphilippe.ovarlez@centralesupelec.fr}).
\and Loong-Fah Cheong \at Department of Electrical and Computer Engineering, National University of Singapore, Singapore (\email{eleclf@nus.edu.sg}).
\and Ali Chehab \at Department of Electrical and Computer Engineering, American University of Beirut, Beirut, Lebanon (\email{chehab@aub.edu.lb}).}

\maketitle

\abstract*{This chapter introduces a novel target detector for hyperspectral imagery. The detector is independent on the unknown covariance matrix, behaves well in large dimensions, distributional free, invariant to atmospheric effects, and does not require any background dictionary to be constructed. Based on a modification of the Robust Principal Component Analysis (RPCA), a given hyperspectral image (HSI) is regarded as being made up of the sum of a low-rank background HSI and a sparse target HSI that contains the targets based on a pre-learned target dictionary specified by the user. The sparse component (that is, the sparse target HSI), will only be the object of interest, and hence, constitutes the novel target detector.}

\abstract{In this work, a novel target detector for hyperspectral imagery is developed. The detector is independent on the unknown covariance matrix, behaves well in large dimensions, distributional free, invariant to atmospheric effects, and does not require a background dictionary to be constructed. Based on a modification of the robust principal component analysis (RPCA), a given hyperspectral image (HSI) is regarded as being made up of the sum of a low-rank background HSI and a sparse target HSI that contains the targets based on a pre-learned target dictionary specified by the user. The sparse component is directly used for the detection, that is, the targets are simply detected at the non-zero entries of the sparse target HSI. Hence, a novel target detector is developed, which is simply a sparse HSI generated automatically from the original HSI, but containing only the targets with the background is suppressed. The detector is evaluated on real experiments, and the results of which demonstrate its effectiveness for hyperspectral target detection especially when the targets are well matched to the surroundings.}

\section{Introduction}
\label{sec:intro}

\subsection{What Is a Hyperspectral Image?}

An airborne hyperspectral imaging sensor is capable of simultaneously acquiring the same spatial scene in a contiguous and multiple narrow (0.01 - 0.02 $\upmu$m) spectral wavelength (color) bands \cite{Shaw02, manolakis2003hyperspectral, manolakis_lockwood_cooley_2016, 974715, 7564440, ZHANG20153102, 6555921, PLAZA2009S110, 6509473, bookJocekynToaPPEAR, DallaMura2011, prasad2011optical, chanussotDec2009}. When all the spectral bands are stacked together, the result is a hyperspectral image (HSI) whose cross-section is a function of the spatial coordinates and its depth is a function of wavelength. Hence, an HSI is a 3-D data cube having two spatial dimensions and one spectral dimension. Thanks to the narrow acquisition, the HSI could have hundreds to thousands of contiguous spectral bands. Having this very high level of spectral detail gives better capability to see the unseen. 

Each band of the HSI corresponds to an image of the surface covered by the field of view of the hyperspectral sensor, whereas each pixel in the HSI is a $p$-dimensional vector, $\mathbf{x}\in\mathbb{R}^p$ ($p$ stands for the total number of spectral bands), consisting of a spectrum characterizing the materials within the pixel. The spectral signature of $ \mathbf{x}$ (also known as reflectance spectrum) shows the fraction of incident energy, typically sunlight, that is reflected by a material from the surface of interest as a function of the wavelength of the energy \cite{manolakis2003hyperspectral, JoanaThesis}.

The HSI usually contains both pure and mixed pixels \cite{manolakis2003hyperspectral, Manolakis09, 6048900, 6504505, 5658102, 6351978}. A pure pixel contains only one single material, whereas a mixed pixel contains multiple materials, with its spectral signature representing the aggregate of all the materials in the corresponding spatial location. The latter situation often arises because hyperspectral images are collected hundreds to thousands of meters away from an object so that the object becomes smaller than the size of a pixel. Other scenarios might involve, for example, a military target hidden under foliage or covered with camouflage material.

\subsection{Hyperspectral Target Detection: Concept and Challenges}

With the rich information afforded by the high spectral dimensionality, hyperspectral imagery has found many applications in various fields, such as astronomy, agriculture \cite{Patel2001, Datt2003}, mineralogy \cite{Lehmann2001}, military \cite{manolakis2002detection, Stein02, 4939406}, and in particular, target detection \cite{Shaw02, manolakis2003hyperspectral, Manolakis14, Manolakis09, manolakis2002detection, 7739987, 7165577, 8069001, Frontera14, 6378408}. Usually, the detection is built using a binary hypothesis test that chooses between the following competing null and alternative hypothesis: target absent ($H_0$), that is, the test pixel $\mathbf{x}$ consists only of background, and target present ($H_1$), where $\mathbf{x}$ may be either fully or partially occupied by the target material. 

It is well known that the signal model for hyperspectral test pixels is fundamentally different from the additive model used in radar and communications applications \cite{manolakis_lockwood_cooley_2016, Manolakis09}. We can regard each test pixel $\mathbf{x}$ as being made up of $\mathbf{x} = \alpha \, \mathbf{t}$ + $(1-\alpha) \, \mathbf{b}$, where $0 \leq \alpha \leq 1$ designates the target fill-fraction, $\mathbf{t}$ is the spectrum of the target, and $\mathbf{b}$ is the spectrum of the background. This model is known as replacement signal model, and hence, when a target is present in a given HSI, it replaces (that is, removes) an equal part of the background \cite{manolakis_lockwood_cooley_2016}. For notational convenience, sensor noise has been incorporated into the target and background spectra (i.e., the vectors $\mathbf{t}$ and $\mathbf{b}$ include noise) \cite{manolakis_lockwood_cooley_2016}. 

In particular, when $\alpha=0$, the pixel $\mathbf{x}$ is fully occupied by the background material (target not present). When $\alpha = 1$, the pixel $\mathbf{x}$ is fully occupied by the target material and is usually referred to as the full or resolved target pixel. Whereas when $0<\alpha<1$, the pixel $\mathbf{x}$ is partially occupied by the target material and is usually referred to as the subpixel or unresolved target \cite{Manolakis09}.

A prior target information can often be provided to the user. In real-world hyperspectral imagery, this prior information may not be only related to its spatial properties (e.g.,  size, shape, texture) and which is usually not at our disposal, but to its spectral signature. The latter usually hinges on the nature of the given HSI where the spectra of the targets of interest have been already measured by some laboratories or with some handheld spectrometers.

Different Gaussian-based target detectors (e.g., Matched Filter \cite{Manolakis00, Nasrabadi08}, Normalized Matched Filter \cite{kraut1999cfar}, and Kelly detector \cite{Kelly86}) have been developed. In these classical detectors, the target of interest to detect is known (e.g., its spectral signature is fully provided to the user).

However, the aforementioned detectors present several limitations in real-world hyperspectral imagery. The task of understanding and solving these limitations presents significant challenges for hyperspectral target detection.

\begin{itemize}
\item{\bf Challenge one}:
One of the major drawbacks of the aforementioned classical target detectors is that they depend on the unknown covariance matrix (of the background surrounding the test pixel) whose entries have to be carefully estimated, especially in large dimensions \cite{LEDOIT2004365, LedoitHoney, AhmadCamsap2017}, and to ensure success under different environments \cite{JoanaThesis, 7165577, 5606730, 6884641, 6894189}. However, estimating large covariance matrices has been a longstanding important problem in many applications and has attracted increased attention over several decades.
When the spectral dimension is considered large compared to the sample size (which is the usual case), the traditional covariance estimators are estimated with a lot of errors unless some covariance regularization methods are considered \cite{LEDOIT2004365, LedoitHoney, AhmadCamsap2017}. It implies that the largest or smallest estimated coefficients in the matrix tend to take on extreme values not because this is ``the truth'', but because they contain an extreme amount of error \cite{LEDOIT2004365, LedoitHoney}. This is one of the main reasons why the classical target detectors usually behave poorly in detecting the targets of interest in a given HSI.
\\
In addition, there is always an explicit assumption (specifically, Gaussian) on the statistical distribution characteristics of the observed data. For instance, most materials are treated as Lambertian because their bidirectional reflectance distribution function characterizations are usually not available, but the actual reflection is likely to have both a diffuse component and a specular component. This latter component would result in gross corruption of the data. In addition, spectra from multiple materials are usually assumed to interact according to a linear mixing model; nonlinear mixing effects are not represented and will contribute to another source of noise.
\\
\item{\bf Challenge two}:
The classical target detectors that depend on the target to detect $\mathbf{t}$ use only a single reference spectrum for the target of interest. This may be inadequate since in real-world hyperspectral imagery, various effects that produce variability to the material spectra (e.g., atmospheric conditions, sensor noise, material composition, and scene geometry) are inevitable \cite{803418, 1000320}.
For instance, target signatures are typically measured in laboratories or in the field with handheld spectrometers that are at most a few inches from the target surface. Hyperspectral images, however, are collected at huge distances away from the target and have significant atmospheric effects present.
\end{itemize}
Recent years have witnessed a growing interest in the notion of sparsity as a way to model signals. The basic assumption of this model is that natural signals can be represented as a ``sparse'' linear combination of atom signals taken from a dictionary. In this regard, two main issues need to be addressed: (1) how to represent a signal in the sparsest way, for a given dictionary? and (2) how to construct an accurate dictionary in order to successfully represent the signal?

 Recently, a signal classification technique via sparse representation was developed for the application of face recognition \cite{4483511}. It is observed that aligned faces of the same object with varying lighting conditions approximately lie in a low-dimensional subspace \cite{1177153}. Hence, a test face image can be sparsely represented by atom signals from all classes. This representation approach has also been exploited in several other signal classification problems such as iris recognition \cite{5339067}, tumor classification \cite{1177153asf}, and hyperspectral imagery unmixing \cite{ 6555921, doi818245, 6200362}. 

In this context, Chen {et al.} \cite{Chen11} have been inspired by the work in \cite{4483511}, and developed an approach for sparse representation of hyperspectral test pixels. In particular, each test pixel $\mathbf{x} \in\mathbb{R}^p$ (either target or background) in a given HSI, is assumed to lie in a low-dimensional subspace of the $p$-dimensional spectral-measurement space. Hence, it can be represented by a very few atom signals taken from the dictionaries, and the recovered sparse representation can be used directly for the detection. For example, if a test pixel $\mathbf{x}$ contains the target (that is, $\mathbf{x} = \alpha \, \mathbf{t} + (1-\alpha)\, \mathbf{b}$, with $0<\alpha \leq 1$), it can be sparsely represented by atom signals taken from the target dictionary (denoted as $\mathbf{A}_t)$; whereas, if $\mathbf{x}$ is only a background pixel (e.g., $\alpha=0$), it can be sparsely represented by atom signals taken from the  background dictionary (denoted as $\mathbf{A}_b$). 
Very recently, Zhang {et al.} \cite{Zhang15} have extended the work done by Chen {\it et al.} in \cite{Chen11} by combining the idea of binary hypothesis and sparse representation together, obtaining a more complete and realistic sparsity model than in \cite{Chen11}. More precisely, Zhang {et al.} \cite{Zhang15} have assumed that if the test pixel $\mathbf{x}$ belongs to hypothesis $H_0$ (target absent), it will be modeled by the $\mathbf{A}_b$ only; otherwise, it will be modeled by the union of $\mathbf{A}_b$ and $\mathbf{A}_t$. This in fact yields a competition between the two hypotheses corresponding to the different pixel class label.

These sparse representation methods \cite{Chen11, Zhang15} are independent on the unknown covariance matrix, behave well in large dimensions, distributional free, and invariant to atmospheric effects. More precisely, they can alleviate the spectral variability caused by atmospheric effects, and can also better deal with a greater range of noise phenomena. 
\begin{itemize}
\item{\bf Challenge three}:
The main drawback of these sparse representation methods \cite{Chen11, Zhang15} is the lack of a sufficiently universal dictionary, especially for the background $\mathbf{A}_b$; some form of in-scene adaptation would be desirable. The background dictionary $\mathbf{A}_b$ is usually constructed using an adaptive scheme (a local method) which is  based on a dual concentric window centered on the test pixel, with an inner window region (IWR) centered within an outer window region (OWR), and only the pixels in the OWR will constitute the samples for $\mathbf{A}_b$. Clearly, the dimension of IWR is very important and has a strong impact on the target detection performance since it aims to enclose the targets of interest to be detected. It should be set larger than or equal to the size of all the desired targets of interest in the corresponding HSI, so as to exclude the target pixels from erroneously appearing in $\mathbf{A}_b$. However, information about the target size in the image is usually not at our disposal. It is also very unwieldy to set this size parameter when the target could be of irregular shape (e.g., searching for lost plane parts of a missing aircraft). Another tricky situation is when there are multiple targets in close proximity in the image (e.g., military vehicles in long convoy formation). Hence, the construction of $\mathbf{A}_b$ for the sparse representation methods is a very challenging problem since a contamination of it by the target pixels can potentially affect the target detection performance.
\end{itemize}

\subsection{Goals and Outline}
\label{sec:sub31}
In this work, we handle all the aforementioned challenges by making very little specific assumptions about the background or target \cite{8462257, 8677268}. Based on a modification of the recently developed robust principal component analysis (RPCA) \cite{Candes11}, our method decomposes an input HSI into a background HSI (denoted by $\mathbf{L}$) and a sparse target HSI (denoted by $\mathbf{E}$) that contains the targets of interest. 

While we do not need to make assumptions about the size, shape, or number of the targets, our method is subject to certain generic constraints that make less specific assumption on the background or the target. These constraints are similar to those used in RPCA \cite{Candes11, NIPS2009_3704}, including
\begin{enumerate}
\item The background is not too heavily cluttered with many different materials with multiple spectra, so that the background signals should span a low-dimensional subspace, a property that can be expressed as the low-rank condition of a suitably formulated matrix \cite{ChenYu, Zhang15b, 7322257, 8260545, 8126244, Ahmad2017a, 7312998}; 
\item The total image area of all the target(s) should be small relative to the whole image (i.e., spatially sparse), e.g., several hundred pixels in a million-pixel image, though there is no restriction on the target shape or the proximity between the targets.
\end{enumerate}
Our method also assumes that the target spectra are available to the user and that the atmospheric influence can be accounted for by the target dictionary $\mathbf{A}_t$. This pre-learned target dictionary $\mathbf{A}_t$ is used to cast the general RPCA into a more specific form, specifically, we further factorize the sparse component $\mathbf{E}$ from RPCA into the product of $\mathbf{A}_t$ and a sparse activation matrix $\mathbf{C}$ \cite{8462257}. This modification is essential to disambiguate the true targets from the background.

 \begin{figure}[!tbp]
 \centering
\minipage{0.84\textwidth}
  \includegraphics[width=\linewidth]{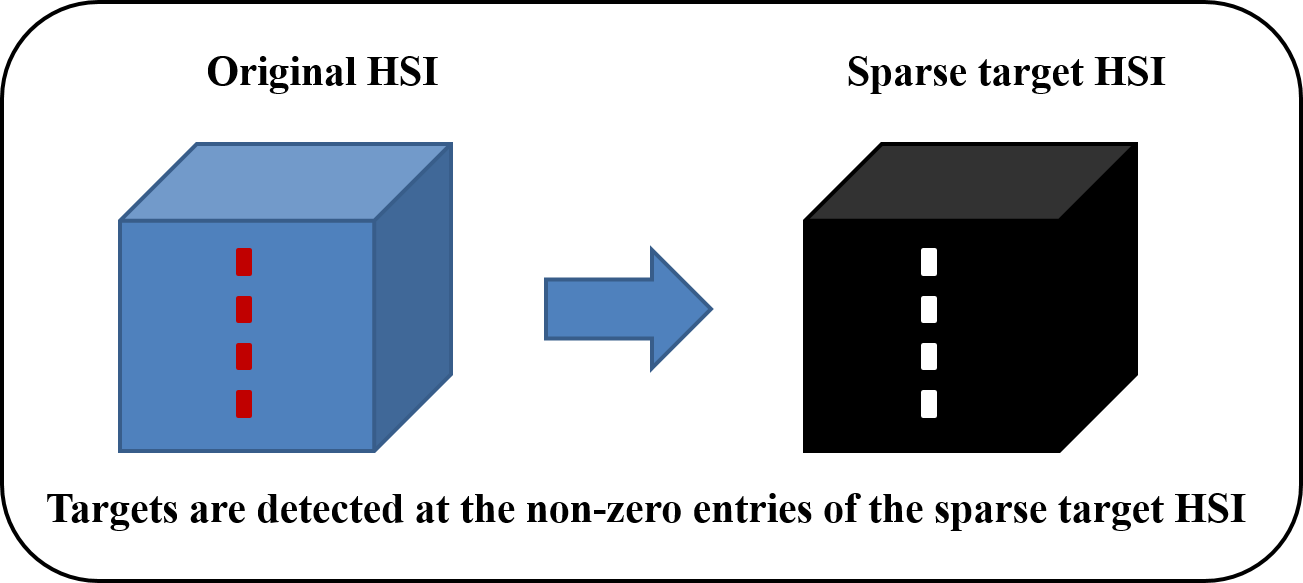}
\endminipage
\caption{Sparse target HSI: our novel target detector.}\label{fig:image1}
\end{figure}
After decomposing a given HSI into the sum of a low-rank HSI and a sparse HSI, the latter will define our detector. That is, the targets are detected at the non-zero entries of the sparse HSI. Hence, a novel target detector is developed, which is simply a sparse HSI generated automatically from the original HSI, but containing only the targets with the background is suppressed (see Fig. \ref{fig:image1}).
\begin{svgraybox}
The main advantages of our proposed detector are the following: (1) independent on the unknown covariance matrix; (2) behaves well in large dimensions; (3) distributional free; (4) invariant to atmospheric effects via the use of the target dictionary $\mathbf{A}_t$; and (5) does not require a background dictionary to be constructed.
\end{svgraybox}
$\\$
This chapter is structured along the following lines. First comes an overview of some related works in Sect. \ref{sec:RelatedWorks}. In Sect. \ref{sec:Maincontribution}, the proposed decomposition model as well as our novel target detector are briefly outlined. Section \ref{sec:experimentss} presents real experiments to gauge the effectiveness of the proposed detector for hyperspectral target detection. The chapter ends with a summary of the work and some directions for future work.

\subsection{Summary of Main Notations} 
Throughout this chapter, we depict vectors in lowercase boldface letters and matrices in uppercase boldface letters. The notation $(.)^T$ and $\mathrm{Tr}(.)$ stand for the transpose and trace of a matrix, respectively. In addition, $\mathrm{rank}(.)$ is for the rank of a matrix. A variety of norms on matrices will be used. For instance, $\mathbf{M}$ is a matrix, and $[\mathbf{M}]_{:,j}$ is the $j$th column. The matrix $l_{2,0}$, $l_{2,1}$ norms are defined by $\left\Vert\mathbf{M}\right\Vert_{2,0} = \# \left\{ j \, : \, \left\Vert\left[\mathbf{M}\right]_{:,j}\right\Vert_2 \, \not= \, 0\right\}$, and $\left\Vert\mathbf{M}\right\Vert_{2,1} = \sum_{j} \left\Vert\left[\mathbf{M}\right]_{:,j}\right\Vert_2$, respectively. The Frobenius norm and the nuclear norm (the sum of singular values of a matrix) are denoted by $\left\Vert\mathbf{M}\right\Vert_F$ and $\left\Vert\mathbf{M}\right\Vert_* = \Tr\left(\mathbf{M}^T \,\mathbf{M}\right)^{(1/2)}$, respectively.

\section{Related Works}
\label{sec:RelatedWorks}
Whatever the real application may be, somehow the general RPCA model needs to be subject to further assumptions for successfully distinguishing the true targets from the background. 
Besides the generic RPCA and our suggested modification discussed in Sect. \ref{sec:sub31}, there have been other modifications of RPCA.
For example, the generalized model of RPCA, named the low-rank representation (LRR) \cite{6180173}, allows the use of a subspace basis as a dictionary or just uses self-representation to obtain the LRR. The major drawback in LRR is that the incorporated dictionary has to be constructed from the background and to be pure from the target samples. This challenge is similar to the aforementioned background dictionary $\mathbf{A}_b$  construction problem. If we use the self-representation form of LRR, the presence of a target in the input image may only bring about a small increase in rank and thus be retained in the background \cite{8677268}.

In the earliest models using a low-rank matrix to represent the background \cite{Candes11, NIPS2009_3704, SPCPCandes}, no prior knowledge on the target was considered. In some applications such as Speech enhancement and hyperspectral imagery, we may expect some prior information about the target of interest and which can be provided to the user. Incorporating this information about the target into the separation scheme in the general RPCA model should allow us to potentially improve the target extraction performance. For example, Chen and Ellis \cite{6701883}, and Sun and Qin \cite{7740039}, proposed a Speech enhancement system by exploiting the knowledge about the likely form of the targeted speech. This was accomplished by factorizing the sparse component from RPCA into the product of a dictionary of target speech templates and a sparse activation matrix. The proposed methods in \cite{6701883} and \cite{7740039} typically differ on how the fixed target dictionary of speech spectral templates is constructed.
Our proposed model in Sect. \ref{sec:Maincontribution} is very related to \cite{6701883} and \cite{7740039}. In real-world hyperspectral imagery, the prior target information may not be only related to its spatial properties (e.g.,  size, shape, and texture) and which is usually not at our disposal, but to its spectral signature. The latter usually hinges on the nature of the given HSI where the spectra of the targets of interest present have been already measured by some laboratories or with some handheld spectrometers.

\section{Main Contribution}
\label{sec:Maincontribution}

Suppose an HSI of size $h \times w \times p$, where $h$ and $w$ are the height and width of the image scene, respectively, and $p$ is the number of spectral bands. Our proposed modification of RPCA is mainly based on the following steps:
\begin{enumerate}
\item Let us consider that the given HSI contains $q$ pixels $\{ \mathbf{x}_i\}_{i\in[1,\,q]}$ of the form:
\begin{align*}
\mathbf{x}_i = \alpha_i \, \mathbf{t}_i + (1 - \alpha_i) \, \mathbf{b}_i, ~~~~0<\alpha_i \leq 1\,,
\end{align*}
where $\mathbf{t}_i$ represents the known target that replaces a fraction $\alpha_i$ of the background $\mathbf{b}_i$ (i.e., at the same spatial location). The remaining ($e-q$) pixels in the given HSI, with $e = h \times w$, are thus only background ($\alpha=0$).
$\\$
\item We assume that all $\{\mathbf{t}_i\}_{i\in[1,\,q]}$ consist of similar materials, and thus they should be represented by a linear combination of $N_t$ common target samples $\{\mathbf{a}^t_{j}\}_{j \in [1, \, N_t]}$, where $\mathbf{a}_j^t \in\mathbb{R}^p$ (the superscript $t$ is for target), but weighted with different set of coefficients $\{\beta_{i,j}\}_{j\in[1, N_t]}$. Thus, each of the $q$ targets is represented as
\begin{align*}
\mathbf{x}_i = \alpha_i \sum\limits_{j=1}^{N_t} \Big(\beta_{i,j} \, \mathbf{a}^t_{j} \Big) + (1 - \alpha_i) \, \mathbf{b}_i \hspace{0.5cm} i \in [1,q]\, . 
\end{align*}
\item We rearrange the given HSI into a two-dimensional matrix $\mathbf{D}\in \mathbb{R}^{e \times p}$, with $e = h \times w$ (by lexicographically ordering the columns). The matrix $\mathbf{D}$, can be decomposed into a low-rank matrix $\mathbf{L}_0$ representing the pure background, a sparse matrix capturing any spatially small signal residing in the known target subspace, and a noise matrix $\mathbf{N}_0$. More precisely, the model is
\begin{align*}\label{eq:mod2}
\small
\mathbf{D} &= \mathbf{L}_0 + (\mathbf{A}_t \, \mathbf{C}_0)^T + \mathbf{N}_0 \, ,
\end{align*}
where $(\mathbf{A}_t\mathbf{C}_0)^T$ is the sparse target matrix, ideally with $q$ non-zero rows representing $\alpha_i \{\mathbf{t}^T_i\}_{i\in[1,q]}$ , with target dictionary $\mathbf{A}_t \in \mathbb{R}^{p \times N_t}$ having columns representing the target samples $\{\mathbf{a}^t_{j}\}_{j \in [1, N_t]}$, and a coefficient matrix $\mathbf{C}_0\in\mathbb{R}^{N_t \times e}$ that should be a sparse column matrix, again ideally containing $q$ non-zero columns each representing $\alpha_i [\beta_{i,1}, \, \ldots, \, \beta_{i, N_t}]^T$, $i \in [1, q]$. $\mathbf{N}_0$ is assumed to be independent and identically distributed Gaussian noise with zero mean and unknown standard deviation.
$\\$
\item After reshaping $\mathbf{L}_0$, $\left(\mathbf{A}_t\, \mathbf{C}_0\right)^T$, and $\mathbf{N}_0$ back to a cube of size $h \times w \times p$, we call these entities as ``low-rank background HSI'', ``sparse target HSI'', and ``noise HSI'', respectively.
\end{enumerate}
In order to recover the low-rank matrix $\mathbf{L}_0$ and the sparse target matrix $\left(\mathbf{A}_t\mathbf{C}_0\right)^T$, we consider the following minimization problem:
\vspace{-0.8mm}
\begin{equation}
\label{eq:ert5}
\underset{\mathbf{L}, \mathbf{C}} {\mathrm{min}} \,  \left\{\tau \, \mathrm{rank}(\mathbf{L})+ \lambda \, \left\Vert\mathbf{C}\right\Vert_{2,0} +  \left\Vert\mathbf{D} - \mathbf{L} - \left(\mathbf{A}_t\mathbf{C}\right)^T\right\Vert_F^2 \right\}\,,
\end{equation}
where $\tau$ controls the rank of $\mathbf{L}$, and $\lambda$ the sparsity level in $\mathbf{C}$.

\subsection{Recovering a Low-Rank Background Matrix and a Sparse Target Matrix by Convex Optimization}
We relax the rank term and the $||.||_{2,0}$ term to their convex proxies. More precisely, we use the nuclear norm $||\mathbf{L}||_*$  as a surrogate for the rank$(\mathbf{L})$ term, and the $l_{2,1}$ norm for the $l_{2,0}$ norm.\footnote{A natural suggestion could be that the rank of $\mathbf{L}$ usually has a physical meaning (e.g., number of endmembers in background), and thus, why not to minimize the latter two terms in Eq. \eqref{eq:convex_model} with the constraint that the rank of $\mathbf{L}$ should not be larger than a fixed value $d$? That is,
\begin{center}
$\underset{\mathbf{L}, \mathbf{C}} {\mathrm{min}} \, \left\{\lambda \,\left\Vert\mathbf{C}\right\Vert_{2,1} +  \left\Vert\mathbf{D} - \mathbf{L} - \left(\mathbf{A}_t \mathbf{C}\right)^T\right\Vert_F^2 \right\}\,,
~~s.t.~~ \mathrm{rank}(\mathbf{L}) \leq d$.
\end{center}
In our opinion, assuming that the number of endmembers in background is known exactly will be a strong assumption and our work will be less general as a result. One can assume $d$ to be some upper bound, in which case, the suggested formulation is a possible one. However, solving such a problem (with a hard constraint that the rank should not exceed some bound) is in general a NP-hard problem, unless there happens to be some special form in the objective which allows for a tractable solution. Thus, we adopt the soft constraint form with the nuclear norm as a proxy for the rank of $\mathbf{L}$; this is an approximation commonly done in the field and is found to give good solutions in many problems empirically.} 

We now need to solve the following convex minimization problem:
\vspace{-0.8mm}
\begin{align}{\label{eq:convex_model}}
\small
\underset{\mathbf{L}, \mathbf{C}} {\mathrm{min}} \, \left\{\tau \,\left\Vert\mathbf{L}\right\Vert_*+ \lambda \,\left\Vert\mathbf{C}\right\Vert_{2,1} +  \left\Vert\mathbf{D} - \mathbf{L} - \left(\mathbf{A}_t \mathbf{C}\right)^T\right\Vert_F^2 \right\}\,.
\end{align}
Problem \eqref{eq:convex_model} is solved via an alternating minimization of two sub-problems. Specifically, at each iteration $k$,
\begin{subequations}\label{eq:sub}
\footnotesize
\begin{alignat}{2}
\label{eq:sub1a}
\mathbf{L}^{(k)} &= \underset{\mathbf{L}} {\mathrm{argmin}} \, \left\{\left\Vert\mathbf{L} - \left(\mathbf{D} - \left(\mathbf{A}_t \, \mathbf{C}^{(k-1)}\right)^T\right)\right\Vert_F^2 + \tau \,\left\Vert\mathbf{L}\right\Vert_* \, \right\}\,, \\
\label{eq:sub2b}
\mathbf{C}^{(k)} &= \underset{\mathbf{C}} {\mathrm{argmin}} \, \left\{\left\Vert\left(\mathbf{D} - \mathbf{L}^{(k)}\right)^T -  \mathbf{A}_t \,  \mathbf{C}\right\Vert_F^2 + \lambda \,\left\Vert\mathbf{C}\right\Vert_{2,1} \, \right\}.
\end{alignat}
\end{subequations}
The minimization sub-problems \eqref{eq:sub1a}, \eqref{eq:sub2b} are convex and each can be solved optimally.
\\
\\
{\bf Solving sub-problem \eqref{eq:sub1a}:}
we solve sub-problem \eqref{eq:sub1a} via the Singular Value Thresholding operator \cite{SVT2010}. We assume that $\left(\mathbf{D} - \left(\mathbf{A}_t \, \mathbf{C}^{(k-1)}\right)^T \right)$ has a rank equal to $r$.
According to Theorem 2.1 in \cite{SVT2010}, sub-problem \eqref{eq:sub1a} admits the following closed-form solution:
\begin{center}
\noindent \fbox{\parbox{7cm}{%
$~~~~~~~~~~~~\mathbf{L}^{(k)} = D_{\tau/2}\left(\mathbf{D} - \left(\mathbf{A}_t \, \mathbf{C}^{(k-1)}\right)^T \right)$\vspace{0.3cm}\\
$~~~~~~~~~~~~~~~~~~~~= \mathbf{U}^{(k)} \, D_{\tau/2} \left( \mathbf{S}^{(k)} \right) \, \mathbf{V}^{(k)T}$\vspace{0.3cm}\\
$~~~~~~~~~~~~~~~~~~~~= \mathbf{U}^{(k)} \, \diag \left(\left\{ \left(s_t^{(k)} - \frac{\tau}{2}\right)_+ \right\} \right) \, \mathbf{V}^{(k)T}$
}}
\end{center}
where $\mathbf{S}^{(k)} = \diag \left( \left\{s_t^{(k)}\right\}_{1 \leq t \leq r}\right)$, and $D_{\tau/2}(.)$ is the singular value shrinkage operator. 
\\
The matrices $\mathbf{U}^{(k)} \in \mathbb{R}^{e \times r}$, $\mathbf{S}^{(k)} \in \mathbb{R}^{r \times r}$, and $\mathbf{V}^{(k)} \in \mathbb{R}^{p \times r}$ are generated by the singular value decomposition (SVD) of $\left(\mathbf{D} - \left(\mathbf{A}_t \, \mathbf{C}^{(k-1)}\right)^T \right)$.

\begin{proof}
Since the function $\left\{\left\Vert\mathbf{L} - \left(\mathbf{D} - \left(\mathbf{A}_t \, \mathbf{C}^{(k-1)}\right)^T\right)\right\Vert_F^2 + \tau \,\left\Vert\mathbf{L}\right\Vert_* \, \right\}$ is strictly convex, it is easy to see that there exists a unique minimizer, and we thus need to prove that it is equal to $D_{\tau/2}\left(\mathbf{D} - \left(\mathbf{A}_t\, \mathbf{C}^{(k-1)}\right)^T \right)$. Note that to understand how the aforementioned closed-form solution has been obtained, we provide in detail the proof steps that have been given in \cite{SVT2010}. 
 
To do this, let us first find the derivative of sub-problem \eqref{eq:sub1a} w.r.t. $\mathbf{L}$ and set it to zero. We obtain
\begin{equation}
\label{eq:r45t}
\left(\mathbf{D} - \left(\mathbf{A}_t \, \mathbf{C}^{(k-1)}\right)^T \right) - \hat{\mathbf{L}} = \frac{\tau}{2} \, \partial \left\Vert\hat{\mathbf{L}}\right\Vert_* \,,
\end{equation}
where $\partial \left\Vert\hat{\mathbf{L}}\right\Vert_*$ is the set of subgradients of the nuclear norm. Let $\mathbf{U}_L\, \mathbf{S}_L\, \mathbf{V}_L^T$ to be the SVD of $\mathbf{L}$, it is known \cite{4797640, Lewis2003, WATSON199233} that
\begin{equation*}
\partial \left\Vert\mathbf{L}\right\Vert_* = \left\{ \mathbf{U}_L\, \mathbf{V}_L^T + \mathbf{W} \, : \, \mathbf{W}\in\mathbb{R}^{e \times p}, \, \mathbf{U}_L^T\, \mathbf{W} = \mathbf{0}, \, \mathbf{W}\, \mathbf{V}_L = \mathbf{0}, \, \left\Vert\mathbf{W}\right\Vert_2 \leq 1\right\} \, .
\end{equation*} 
Set $\hat{\mathbf{L}} = D_{\tau/2}\left(\mathbf{D} - \left(\mathbf{A}_t \,\mathbf{C}^{(k-1)}\right)^T \right)$ for short. In order to show that $\hat{\mathbf{L}}$ obeys Eq. \eqref{eq:r45t}, suppose the SVD of $\left(\mathbf{D} - \left(\mathbf{A}_t \,\mathbf{C}^{(k-1)}\right)^T \right)$ is given by
\begin{equation*}
\left(\mathbf{D} - \left(\mathbf{A}_t \,\mathbf{C}^{(k-1)}\right)^T \right) = \mathbf{U}_0 \, \mathbf{S}_0\,\mathbf{V}_0^T + \mathbf{U}_1\,\mathbf{S}_1\,\mathbf{V}_1^T \, ,
\end{equation*}
where $\mathbf{U}_0$, $\mathbf{V}_0$ (resp. $\mathbf{U}_1$, $\mathbf{V}_1$) are the singular vectors associated with singular values larger than $\tau/2$ (resp. inferior than or equal to $\tau/2$).
With these notations, we have
\begin{equation*}
\hat{\mathbf{L}} =  D_{\tau/2}\left(\mathbf{U}_0 \,\mathbf{S}_0 \,\mathbf{V}_0^T \right) = \left(\mathbf{U}_0 \, \left(\mathbf{S}_0 - \frac{\tau}{2}\, \mathbf{I}\right)\, \mathbf{V}_0^T \right) \,.
\end{equation*}
Thus, if we return back to Eq. \eqref{eq:r45t}, we obtain
\begin{align*}
\mathbf{U}_0 \, \mathbf{S}_0\,\mathbf{V}_0^T + \mathbf{U}_1\,\mathbf{S}_1\,\mathbf{V}_1^T - \mathbf{U}_0\, \left(\mathbf{S}_0 - \frac{\tau}{2}\, \mathbf{I}\right)\,\mathbf{V}_0^T = \frac{\tau}{2} \, \partial \left\Vert\hat{\mathbf{L}}\right\Vert_* \, ,
\\
\Rightarrow \mathbf{U}_1\, \mathbf{S}_1\,\mathbf{V}_1^T + \mathbf{U}_0 \,\frac{\tau}{2}\,\mathbf{V}_0^T = \frac{\tau}{2} \, \partial \left\Vert\hat{\mathbf{L}}\right\Vert_* \, ,
\\
\Rightarrow \left(\mathbf{U}_0\, \mathbf{V}_0^T + \mathbf{W}\right) = \partial \left\Vert\hat{\mathbf{L}}\right\Vert_* \, ,
\end{align*}
where $\mathbf{W} = \displaystyle \frac{2}{\tau} \, \mathbf{U}_1\, \mathbf{S}_1\, \mathbf{V}_1^T$. \\

By definition, $\mathbf{U}_0^T\, \mathbf{W} = \mathbf{0}$, $\mathbf{W}\, \mathbf{V}_0 = \mathbf{0}$, and we also have $\left\Vert\mathbf{W}\right\Vert_2 \leq 1$.
\\
Hence, $\left(\mathbf{D} - \left(\mathbf{A}_t \, \mathbf{C}^{(k-1)}\right)^T \right) - \hat{\mathbf{L}} = \displaystyle\frac{\tau}{2} \, \partial \left\Vert\hat{\mathbf{L}}\right\Vert_*$, which concludes the proof.
\end{proof}
$\\$

{\bf Solving sub-problem \eqref{eq:sub2b}:}
\eqref{eq:sub2b} can be solved by various methods, among which we adopt the alternating direction method of multipliers (ADMM) \cite{Boyd:2011:DOS:2185815.2185816}. 
More precisely, we introduce an auxiliary variable $\mathbf{F}$ into sub-problem \eqref{eq:sub2b} and recast it into the following form:
\begin{equation}
\label{eq:problemm}
\left(\mathbf{C}^{(k)}, \mathbf{F}^{(k)}\right) = \underset{s.t.~~ \mathbf{C} = \mathbf{F}} {\mathrm{argmin}} \, \left\{\left\Vert \left(\mathbf{D} - \mathbf{L}^{(k)}\right)^T - \mathbf{A}_t \, \mathbf{C}\right\Vert_F^2 + \lambda \, \left\Vert\mathbf{F}\right\Vert_{2,1}\right\} \,.
\end{equation}
Problem \eqref{eq:problemm} is then solved as follows (scaled form of ADMM):
\begin{subequations}\label{eq:subbb}
\small
\begin{alignat}{2}
\label{eq:sub11a}
\mathbf{C}^{(k)} &= \underset{\mathbf{C}} {\mathrm{argmin}} \, \left\{\left\Vert \left(\mathbf{D} - \mathbf{L}^{(k)}\right)^T - \mathbf{A}_t \,\mathbf{C}\right\Vert_F^2 + \frac{\rho^{(k-1)}}{2}\, \left\Vert\mathbf{C} - \mathbf{F}^{(k-1)} + \frac{1}{\rho^{(k-1)}}\, \mathbf{Z}^{(k-1)}\right\Vert_F^2 \, \right\}\,, \\
\label{eq:sub22b}
\mathbf{F}^{(k)} &= \underset{\mathbf{F}} {\mathrm{argmin}} \, \left\{\lambda \, \left\Vert\mathbf{F}\right\Vert_{2,1} + \frac{\rho^{(k-1)}}{2}\,\left\Vert \mathbf{C}^{(k)} - \mathbf{F} + \frac{1}{\rho^{(k-1)}} \, \mathbf{Z}^{(k-1)}\right\Vert_F^2\right\}\,, \\
\label{eq:sub33c}
\mathbf{Z}^{(k)} &= \mathbf{Z}^{(k-1)} + \rho^{(k-1)} \,\left(\mathbf{C}^{(k)} - \mathbf{F}^{(k)}\right)\,, 
\end{alignat}
\end{subequations}
where $\mathbf{Z} \in \mathbb{R}^{N_t \times e}$ is the Lagrangian multiplier matrix, and $\rho$ is a positive scalar.
\\
\\
{\em Solving sub-problem \eqref{eq:sub11a}}:
\begin{align*}
-2\, \mathbf{A}_t^T \left(\left(\mathbf{D} - \mathbf{L}^{(k)}\right)^T - \mathbf{A}_t\, \mathbf{C} \right) + \rho^{(k-1)} \, \left( \mathbf{C} - \mathbf{F}^{(k-1)} + \frac{1}{\rho^{(k-1)}} \, \mathbf{Z}^{(k-1)}\right) = \mathbf{0} \, ,
\\
\Rightarrow \left(2\, \mathbf{A}_t^T \, \mathbf{A}_t + \rho^{(k-1)}\, \mathbf{I} \right) \, \mathbf{C} = \rho^{(k-1)} \, \mathbf{F}^{(k-1)} - \mathbf{Z}^{(k-1)} + 2\, \mathbf{A}_t^T \, \left(\mathbf{D} - \mathbf{L}^{(k)}\right)^T \, . 
\end{align*}
This implies
\[
 \boxed{\mathbf{C}^{(k)} = \left( 2\, \mathbf{A}_t^T \, \mathbf{A}_t + \rho^{(k-1)} \, \mathbf{I}\right)^{-1} \, \left(\rho^{(k-1)}\, \mathbf{F}^{(k-1)} - \mathbf{Z}^{(k-1)} + 2\, \mathbf{A}_t^T \, \left(\mathbf{D} - \mathbf{L}^{(k)}\right)^T \right)}
 \] 
\\
\\
{\em Solving sub-problem \eqref{eq:sub22b}}:
\\
\\
According to Lemma 3.3 in \cite{doi:10.1137/080730421} and Lemma 4.1 in \cite{6180173}, sub-problem \eqref{eq:sub22b} admits the following closed-form solution:
\[
 \boxed{[\mathbf{F}]_{:, j}^{(k)}  = \max \left( \left\Vert\left[\mathbf{C}\right]_{:,j}^{(k)} + \frac{1}{\rho^{(k-1)}} \, \left[\mathbf{Z}\right]_{:,j}^{(k-1)}\right\Vert_2 - \frac{\lambda}{\rho^{(k-1)}}, \, 0 \right) \, \left( \displaystyle \frac{\left[\mathbf{C}\right]_{:,j}^{(k)} + 
 {\frac{1}{\rho^{(k-1)}}}\, \left[\mathbf{Z}\right]_{:,j}^{(k-1)}}{\left\Vert  \left[\mathbf{C}\right]_{:,j}^{(k)} + \frac{1}{\rho^{(k-1)}}\, \left[\mathbf{Z}\right]_{:,j}^{(k-1)}\right\Vert_2}\right)
 }
 \]
 \\
 \begin{proof}
 At the $j${th} column, sub-problem \eqref{eq:sub22b} refers to
 \begin{equation*}
 \left[\mathbf{F}\right]_{:,j}^{(k)} = \underset{ \left[\mathbf{F}\right]_{:,j}} {\mathrm{argmin}} \, \left\{\lambda \, \left\Vert \left[\mathbf{F}\right]_{:,j}\right\Vert_2 + \frac{\rho^{(k-1)}}{2}\,\left\Vert\left[\mathbf{C}\right]_{:,j}^{(k)} - \left[\mathbf{F}\right]_{:,j} + \frac{1}{\rho^{(k-1)}}\, \left[\mathbf{Z}\right]_{:,j}^{(k-1)}\right\Vert_2^2 \, \right\} \, .
 \end{equation*}
 By finding the derivative w.r.t $[\mathbf{F}]_{:,j}$ and setting it to zero, we obtain
\begin{eqnarray}
 \label{eq:huyrz}
 -\rho^{(k-1)} \, \left( \left[\mathbf{C}\right]_{:,j}^{(k)} -  \left[\mathbf{F}\right]_{:,j} + \frac{1}{\rho^{(k-1)}}\, \left[\mathbf{Z}\right]_{:,j}^{(k-1)} \right) + \frac{\lambda \, \left[\mathbf{F}\right]_{:,j}}{\left\Vert\left[\mathbf{F}\right]_{:,j}\right\Vert_2} = \mathbf{0}\, \nonumber
 \\
 \Rightarrow \left[\mathbf{C}\right]_{:,j}^{(k)} + \frac{1}{\rho^{(k-1)}}\, \left[\mathbf{Z}\right]_{:,j}^{(k-1)} = \left[\mathbf{F}\right]_{:,j} + \frac{\lambda \, \left[\mathbf{F}\right]_{:,j}}{\rho^{(k-1)}\, \left\Vert\left[\mathbf{F}\right]_{:,j}\right\Vert_2} \, .
\end{eqnarray}
 By computing the $l_2$ norm of \eqref{eq:huyrz}, we obtain
 \begin{equation}
  \label{eq:bbcv}
 \left\Vert \left[\mathbf{C}\right]_{:,j}^{(k)} + \frac{1}{\rho^{(k-1)}}\, \left[\mathbf{Z}\right]_{:,j}^{(k-1)}\right\Vert_2 = \left\Vert\left[\mathbf{F}\right]_{:,j}\right\Vert_2 + \frac{\lambda}{\rho^{(k-1)}} \, .
 \end{equation}
From Eq. \eqref{eq:huyrz} and \eqref{eq:bbcv}, we have
\begin{equation}
\label{eq:wwwe}
\displaystyle \frac{\left[\mathbf{C}\right]_{:,j}^{(k)} + \displaystyle\frac{1}{\rho^{(k-1)}}\, \left[\mathbf{Z}\right]_{:,j}^{(k-1)}}{ \left\Vert\left[\mathbf{C}\right]_{:,j}^{(k)} + \displaystyle\frac{1}{\rho^{(k-1)}}\, \left[\mathbf{Z}\right]_{:,j}^{(k-1)}\right\Vert_2} = \frac{\left[\mathbf{F}\right]_{:,j}}{\left\Vert\left[\mathbf{F}\right]_{:,j}\right\Vert_2} \, .
\end{equation}
Consider that
\begin{equation}
\label{eq:heqrtqr}
[\mathbf{F}]_{:,j}=  \left\Vert[\mathbf{F}]_{:,j}\right\Vert_2 \times \displaystyle \frac{[\mathbf{F}]_{:,j}}{ \left\Vert[\mathbf{F}]_{:,j}\right\Vert_2} \, .
\end{equation}
By replacing $ \left\Vert[\mathbf{F}]_{:,j}\right\Vert_2$ from \eqref{eq:bbcv} into \eqref{eq:heqrtqr}, and $\displaystyle\frac{[\mathbf{F}]_{:,j}}{ \left\Vert[\mathbf{F}]_{:,j}\right\Vert_2}$ from \eqref{eq:wwwe} into \eqref{eq:heqrtqr}, we conclude the proof.
 \end{proof}

 \subsection{Some Initializations and Convergence Criterion}
 We initialize $\mathbf{L}^{(0)} = \boldsymbol{0}$, $\mathbf{F}^{(0)} = \mathbf{C}^{(0)} = \mathbf{Z}^{(0)} = \boldsymbol{0}$, $\rho^{(0)} = 10^{-4}$ and update $\rho^{(k)} = 1.1 \, \rho^{(k-1)}$. The criteria for convergence of sub-problem \eqref{eq:sub2b} is $ \left\Vert\mathbf{C}^{(k)} - \mathbf{F}^{(k)} \right\Vert_F^2   \leq 10^{-6}$.

For Problem \eqref{eq:convex_model}, we stop the iteration when the following convergence criterion is satisfied:
\begin{align*}
\frac{ \left\Vert\mathbf{L}^{(k)} - \mathbf{L}^{(k-1)} \right\Vert_F}{ \left\Vert\mathbf{D}\right\Vert_F} \leq \epsilon~~~~~\text{and} ~~~~~
\frac{ \left\Vert\left(\mathbf{A}_t \, \mathbf{C}^{(k)}\right)^T - \left(\mathbf{A}_t \, \mathbf{C}^{(k-1)}\right)^T\right\Vert_F}{ \left\Vert\mathbf{D}\right\Vert_F} \leq \epsilon
\end{align*}
where $\epsilon>0$ is a precision tolerance parameter. We set $\epsilon = 10^{-4}$.

\subsection{Our Novel Target Detector: $(\mathbf{A}_t\mathbf{C})^T$}
\label{sec:first_dete_strat}
We use $(\mathbf{A}_t \mathbf{C})^T$ directly for the detection. \begin{svgraybox}Note that for this detector, we require as few false alarms as possible to be deposited in the target image, but we do not need the target fraction to be entirely removed from the background (that is, a very weak target separation can suffice). As long as enough of the target fractions are moved to the target image, such that non-zero support is detected at the corresponding pixel location, it will be adequate for our detection scheme. From this standpoint, we should choose a $\lambda$ value that is relatively large so that the target image is really sparse with zero or little false alarms, and only the signals that reside in the target subspace specified by $\mathbf{A}_t$ will be deposited there.\end{svgraybox}

\section{Experiments and Analysis}
\label{sec:experimentss}
\begin{figure}[!tbp]
\centering
\minipage{2.5\textwidth}
  \hspace{-1.6cm}\includegraphics[width=0.5\linewidth]{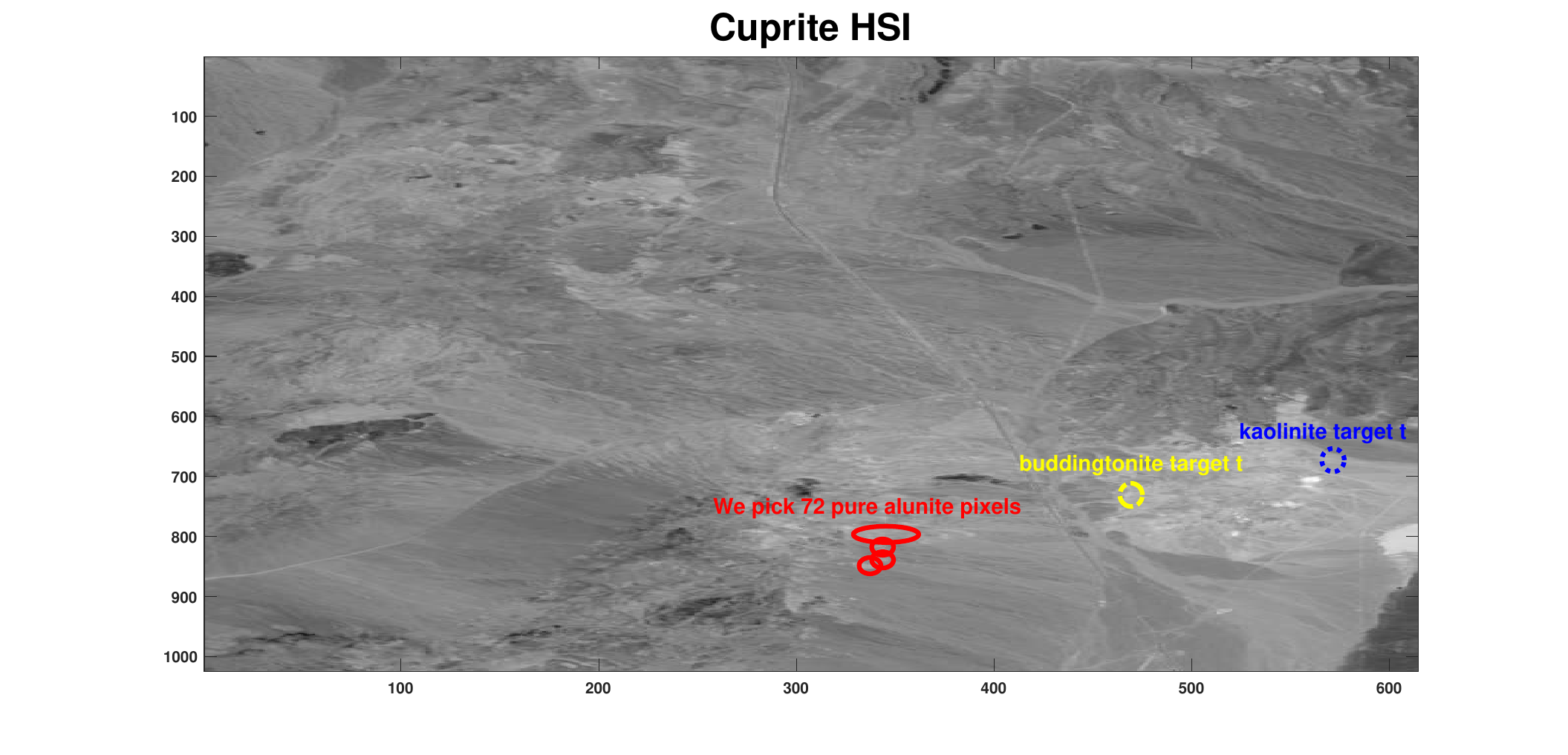}
\endminipage
\caption{The Cuprite HSI of size $1024 \times 614 \times 186$. We exhibit the mean power in db over the 186 spectral bands.}\label{fig:cupriteHSI}
\end{figure}

To obtain the same scene as in Fig. 8 in \cite{Swayze10245}, we have concatenated two sectors labeled as  ``f970619t01p02\_r02\_sc03.a.rf'' and ``f970619t01p02\_r02\_sc04.a.rfl'' from the online Cuprite data \cite{CupriteHSIOnline}. We shall call the resulting HSI as ``Cuprite HSI'' (see Fig. \ref{fig:cupriteHSI}). The Cuprite HSI is a mining district area, which is well understood mineralogically \cite{Swayze10245, JGRE:JGRE1642}. It contains well-exposed zones of advanced argillic alteration, consisting principally of kaolinite, alunite, and hydrothermal silica. It was acquired by the Airborne Visible/Infrared Imaging Spectrometer (AVIRIS) in June 23, 1995 at local noon and under high visibility conditions by a NASAER-2 aircraft flying at an altitude of 20 km. It is a $1024 \times 614$ image and consists of 224 spectral (color) bands in contiguous (of about 0.01 $\upmu$m) wavelengths ranging exactly from 0.4046 to 2.4573 $up\mu$m. Prior to some analysis of the Cuprite HSI, the spectral bands 1-4, 104-113, 148-167, and 221-224 are removed due to the water absorption in those bands. As a result, a total of 186 bands are used.\footnote{We regret that in our work in \cite{8462257, 8677268}, we missed to add ``221-224'' with the other bands that are removed. Adding ``221-224'' will give exactly a total of 186 bands.}
 \begin{figure}[!tbp]
\minipage{0.50\textwidth}
  \includegraphics[width=\linewidth]{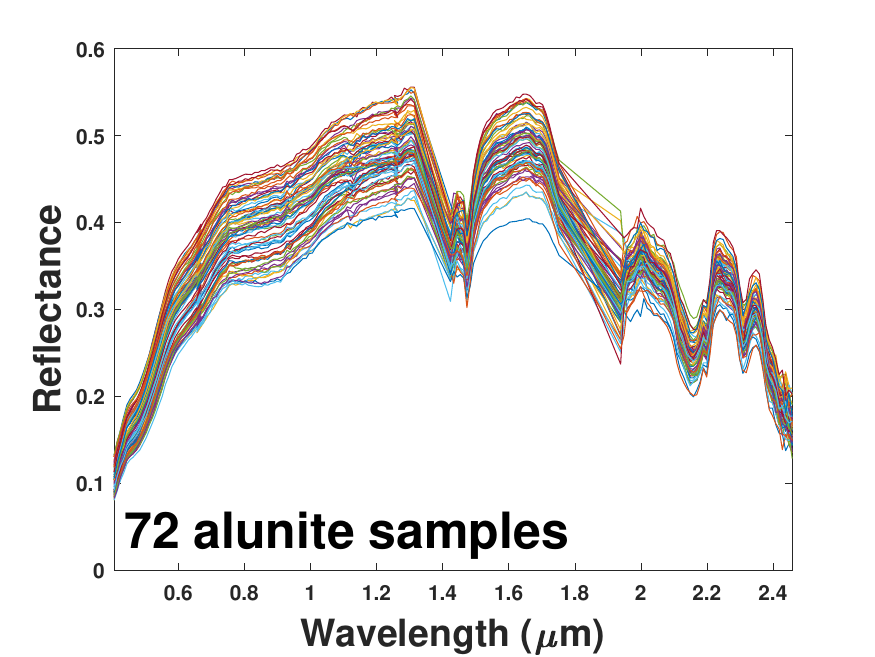}
\endminipage\hfill
\minipage{0.50\textwidth}
  \includegraphics[width=\linewidth]{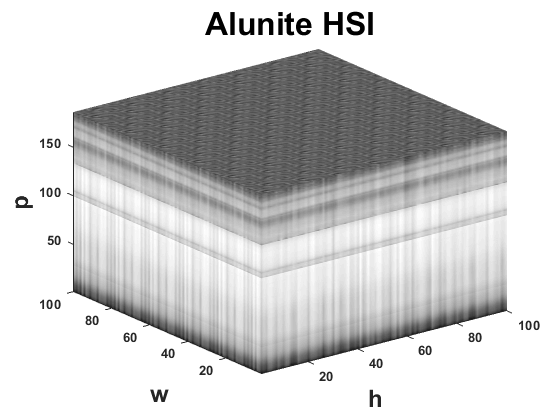}
\endminipage\hfill
\centering
\minipage{0.50\textwidth}
 \includegraphics[width=\linewidth]{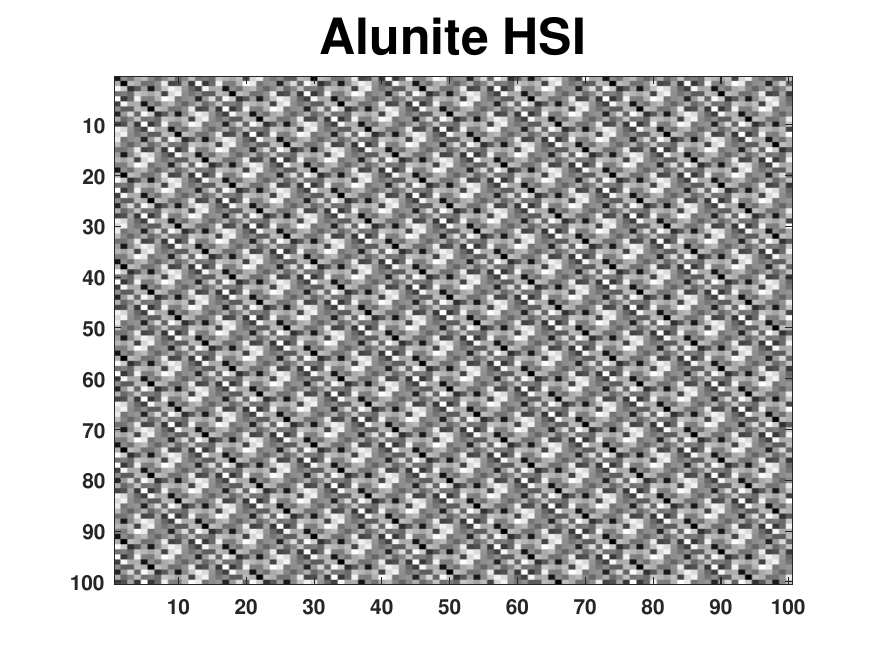}
\endminipage
\caption{A $100 \times 100 \times 186$ ``Alunite HSI'' generated by 72 pure alunite samples picked from the Cuprite HSI (72 pixels from the solid red ellipses in Fig. \ref{fig:cupriteHSI}). For the third image, we exhibit the mean power in db over the 186 spectral bands.}\label{fig:image2}
\end{figure}

By referring to Fig. 8 in \cite{Swayze10245}, we picked 72 pure alunite pixels from the Cuprite HSI (72 pixels located inside the solid red ellipses in Fig. \ref{fig:cupriteHSI}) and generate a $100 \times 100 \times 186$ HSI zone formed by these pixels.
We shall call this small HSI zone as ``Alunite HSI'' (see Fig. \ref{fig:image2}), and which will be used for the target evaluations later. We incorporate, in this zone, seven target blocks (each of size $6\times 3$) with $\alpha\in[0.01, \, 1]$ (all have the same $\alpha$ value), placed in long convoy formation all formed by the same target $\mathbf{t}$ that we picked from the Cuprite HSI and which will constitute our target of interest to be detected. The target $\mathbf{t}$ replaces a fraction $\alpha\in[0.01, \, 1]$ from the background; specifically, the following values of $\alpha$ are considered: 0.01, 0.02, 0.05, 0.1, 0.3, 0.5, 0.8, and 1.
\\
In the experiments, two kinds of target $\mathbf{t}$ are considered:
\begin{enumerate}
\item `$\mathbf{t}$' that represents the buddingtonite target,
\item `$\mathbf{t}$' that represents the kaolinite target.
\end{enumerate} 
More precisely, our detector $\left(\mathbf{A}_t\mathbf{C}\right)^T$   is evaluated on two target detection scenarios:
\begin{itemize}
\item{\bf Evaluation on an easy target (buddingtonite target)}:  It has been noted by Gregg et al. \cite{Swayze10245} that the ammonia in the Tectosilicate mineral type, known as buddingtonite, has a distinct N-H combination absorption at 2.12 $\upmu$m, a position similar to that of the cellulose absorption in dried vegetation, from which it can be distinguished based on its narrower band width and asymmetry.  Hence, the buddingtonite mineral can be considered as an ``easy target'' 
because it does not look like any other mineral with its distinct 2.12 $\upmu$m absorption (that is, it is easily recognized based on its unique 2.12 $\upmu$m absorption band). 
\\
In the experiments,\footnote{The MATLAB code of the proposed detector and experiments is available upon request. Please feel free to contact Ahmad W. Bitar.} we consider the ``buddingtonite'' pixel at location (731, 469) in the Cuprite HSI (the center of the dash-dotted yellow circle in Fig. \ref{fig:cupriteHSI}) as the buddingtonite target $\mathbf{t}$ to be incorporated in the Alunite HSI for $\alpha\in[0.01, \, 1]$.
\begin{figure}[!]
\minipage{0.52\textwidth}
  \includegraphics[width=\linewidth]{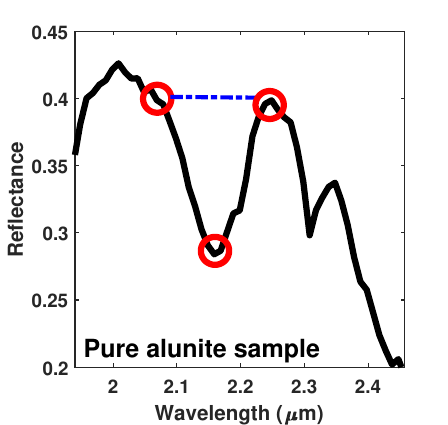}
  \begin{center}
  \bf (a)
  \end{center}
\endminipage\hfill
\minipage{0.52\textwidth}
  \includegraphics[width=\linewidth]{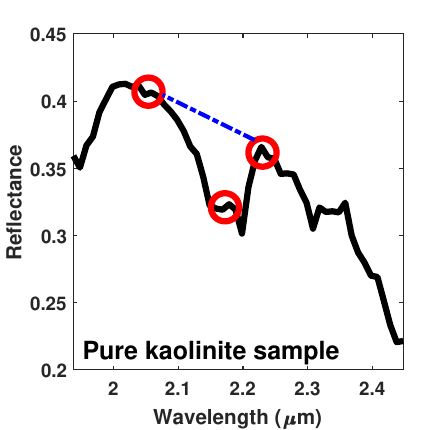}
    \begin{center}
  \bf (b)
  \end{center}
\endminipage
\vspace{-0.3cm}
\caption{Three-point band depth images for both {\bf(a)} alunite and {\bf(b)} kaolinite.}\label{fig:image3}
\end{figure}
$\\$
\item{\bf Evaluation on a challenging target (kaolinite target)}\footnote{We thank Dr. Gregg A. Swayze from the United States Geological Survey (USGS) who has suggested us to evaluate our model \eqref{eq:convex_model} on the distinction between alunite and kaolinite minerals.}{:} The paradigm in military applications for hyperspectral imagery seems to center on finding the target but ignoring all the rest. Sometimes, that rest is important especially if the target is well matched to the surroundings. It has been shown by Gregg et al. \cite{Swayze10245} that both alunite and kaolinite minerals have overlapping spectral features, and thus, discrimination between these two minerals is a big challenge \cite{Swayze10245, doi:10.1029/2002JE001847}.
\\
In the experiments, we consider the ``kaolinite'' pixel at location (672, 572) in the Cuprite HSI (the center of the dotted blue circle in Fig. \ref{fig:cupriteHSI}) as the kaolinite target $\mathbf{t}$ to be incorporated in the Alunite HSI for $\alpha\in[0.01, \, 1]$.  
\\
Figure 4a exhibits a three-point band depth image for our alunite background that shows the locations where an absorption feature, centered near 2.17 $\upmu$m, is expressed in spectra of surface materials. Figure 4b exhibits a three-point band depth image for our kaolinite target that shows the locations where an absorption feature, centered near 2.2 $\upmu$m, is expressed in spectra of surface materials. As we can observe, there is a subtle difference between the alunite and kaolinite three-point band depth images, showing that the successful spectral distinction between these two minerals is a very challenging task to achieve \cite{doi:10.1029/2002JE001847}.\footnote{We have been inspired by Fig. 8D-E in \cite{doi:10.1029/2002JE001847} to provide a close example of it in this chapter as can be shown in Fig. \ref{fig:image3}.}
\end{itemize}

\subsection{Construction of the Target Dictionary $\mathbf{A}_t$}
\label{sec:dict_const}
 \begin{figure}[!tbp]
  \begin{center}
\large \bf When $\mathbf{A}_t$ is constructed from background samples 
\end{center}
\minipage{0.52\textwidth}
  \includegraphics[width=\linewidth]{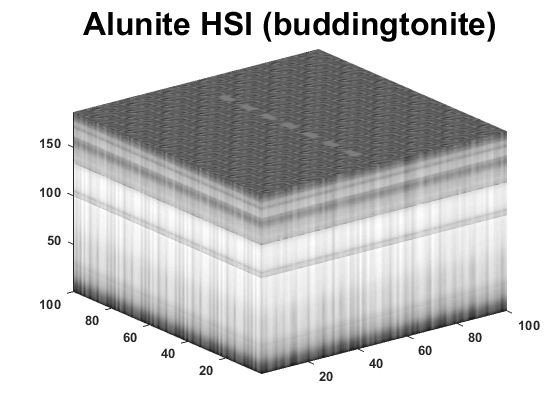}
\endminipage\hfill
\minipage{0.52\textwidth}
  \includegraphics[width=\linewidth]{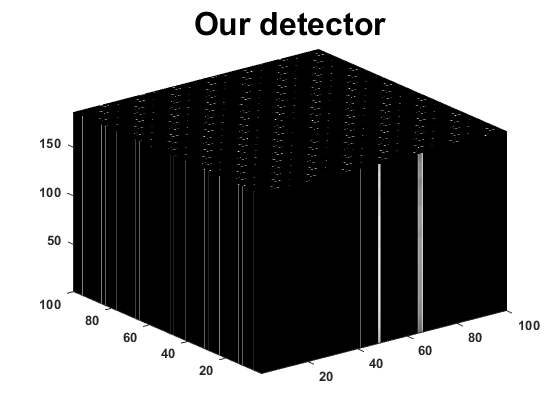}
\endminipage\hfill
\minipage{0.52\textwidth}
  \includegraphics[width=\linewidth]{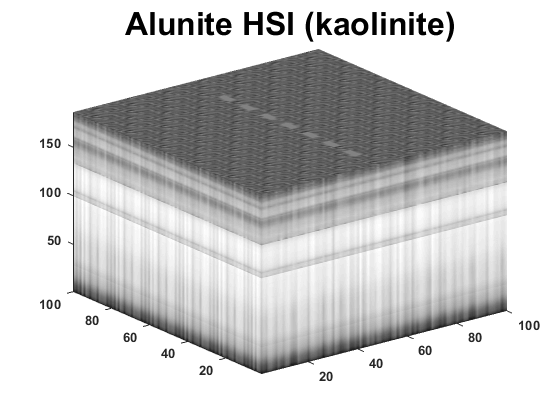}
\endminipage\hfill
\minipage{0.52\textwidth}
  \includegraphics[width=\linewidth]{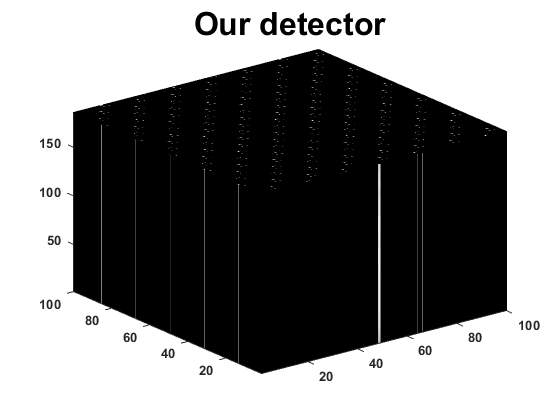}
\endminipage
\caption{Evaluation of our detector $(\mathbf{A}_t\mathbf{C})^T$ for detecting the buddingtonite and kaolinite target (for $\alpha=1$) from the Alunite HSI when $\mathbf{A}_t$ is contsructed from some background pixels acquired from the Alunite HSI.} \label{fig:poor_dictionary}
\end{figure}
An important problem that requires a very careful attention is the construction of an appropriate dictionary $\mathbf{A}_t$ in order to capture the target well and distinguish it from the background. If $\mathbf{A}_t$ does not well represent the target of interest, our model in \eqref{eq:convex_model} may fail on discriminating the targets from the background. For example, Fig. \ref{fig:poor_dictionary} shows the detection results of our detector $(\mathbf{A}_t\mathbf{C})^T$ when $\mathbf{A}_t$ is constructed from some of the background pixels in the Alunite HSI. We can obviously observe that our detector is not able to capture the targets mainly because of the poor dictionary $\mathbf{A}_t$ constructed. 

The target present in the HSI can be highly affected by the atmospheric conditions, sensor noise, material composition, and scene geometry. This may produce huge variations on the target spectra. In view of these real effects, it is very difficult to model the target dictionary $\mathbf{A}_t$ well. But this raises the question on ``{\em how these effects should be dealt with?}''.

Some scenarios for modeling the target dictionary have been suggested in the literature. For example, by using physical models and the MODTRAN atmospheric modeling program \cite{Berk89}, target spectral signatures can be generated under various atmospheric conditions. For simplicity, we handle this problem in this work by exploiting target samples that are available in some online spectral libraries. More precisely, $\mathbf{A}_t$ can be constructed via the United States Geological Survey (USGS - Reston) spectral library \cite{Clark93}. However, the user can also deal with the Advanced Spaceborne Thermal Emission and Reflection (ASTER) spectral library \cite{Baldridge09} that includes data from the USGS spectral library, the Johns Hopkins University (JHU) spectral library, and the Jet Propulsion Laboratory (JPL) spectral library.   

\begin{figure}[!tbp]
 \minipage{0.52\textwidth}
  \includegraphics[width=\linewidth]{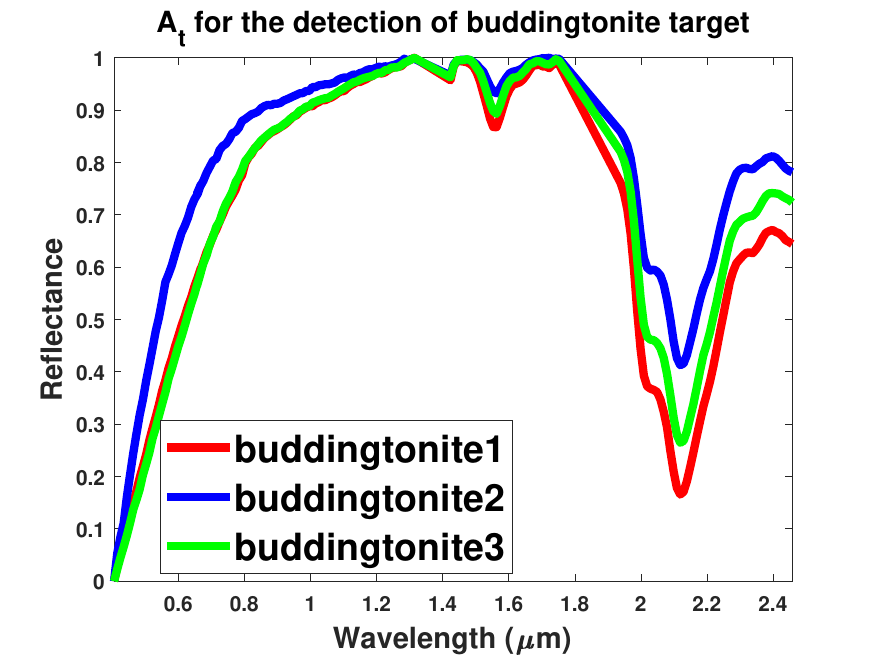}
\endminipage\hfill
\minipage{0.52\textwidth}
  \includegraphics[width=\linewidth]{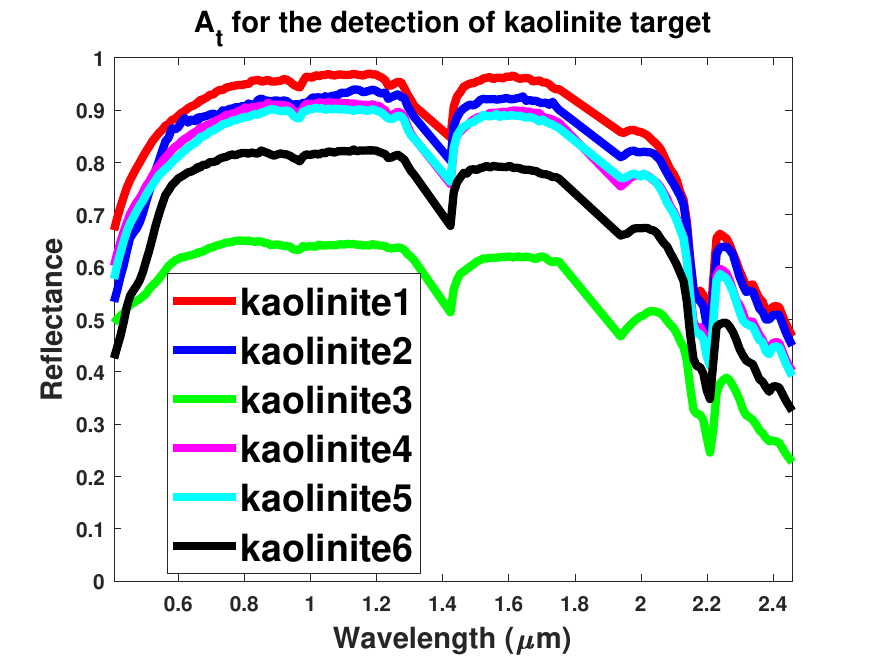}
\endminipage
\caption{Target dictionaries for the detection of buddingtonite and kaolinite.}\label{fig:dictionaries_kao_bud}
\end{figure}

There are three buddingtonite samples available in the ASTER spectral library and will be considered to construct the dictionary $\mathbf{A}_t$ for the detection of our buddingtonite target (see Fig. \ref{fig:dictionaries_kao_bud} (first column)); whereas six kaolinite samples are available in the USGS spectral library and will be acquired to construct $\mathbf{A}_t$ for the detection of our kaolinite target (see Fig. \ref{fig:dictionaries_kao_bud} (second column)). 

Note that the Alunite HSI, the buddingtonite target $\mathbf{t}$, the kaolinite target $\mathbf{t}$, and the buddingtonite/kaolinite target samples extracted from the online spectral libraries are all normalized to the values between 0 and 1.

For instance, it is usually difficult to find, for a specific given target, a sufficient number of available samples in the online spectral libraries. Hence, the dictionary $\mathbf{A}_t$ may still be not sufficiently selective and accurate. This is the most reason why problem \eqref{eq:convex_model} may fail to well capture the targets from the background.

\subsection{Target Detection Evaluation}
We now aim to qualitatively evaluate the target detection performances of our detector $(\mathbf{A}_t\mathbf{C})^T$ on both the buddingtonite and kaolinite target detection scenarios, when $\mathbf{A}_t$ is constructed from target samples that are available in the online spectral libraries (from Fig. \ref{fig:dictionaries_kao_bud}). As can be seen from Fig. \ref{fig:detection_evaluations}, our detector is able to detect the buddingtonite targets with no false alarms until $\alpha \leq 0.1$ where a lot of false alarms appear. 

For the detection of kaolinite, it was difficult to have a clean detection (without false alarms) even for high values of $\alpha$. This is to be expected since the kaolinite target is well matched to the alunite background (the kaolinite and alunite have overlapping spectral features), and hence, the discrimination between them is very challenging.

It is interesting to note (results omitted here) that if we consider $\mathbf{A}_t = \mathbf{t}$ (that is, we are searching for the exact signature $\mathbf{t}$ in the Alunite HSI), the buddingtonite and even the kaolinite targets are able to be detected with no false alarms for $0.1<\alpha\leq1$. When $\alpha\leq0.1$, a lot of false alarms appear, but the detection performances for both the buddingtonite and kaolinite targets remain better than to those in Fig. \ref{fig:detection_evaluations}.

 \begin{figure}[!tbp]
   \begin{center}
\large \bf When $\mathbf{A}_t$ is constructed from target samples 
\end{center}
 \begin{center}
\large$\bf \boldsymbol{\alpha} = 1$
\end{center}
\minipage{0.54\textwidth}
  \includegraphics[width=\linewidth]{originalHSI_alpha1_buddingtonite_cube.png}
\endminipage\hfill
\minipage{0.54\textwidth}
  \includegraphics[width=\linewidth]{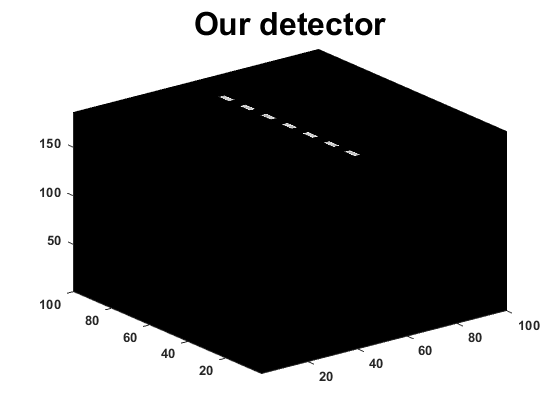}
\endminipage\hfill
\minipage{0.54\textwidth}
  \includegraphics[width=\linewidth]{originalHSI_alpha1_kaolinite_cube.png}
\endminipage\hfill
\minipage{0.54\textwidth}
  \includegraphics[width=\linewidth]{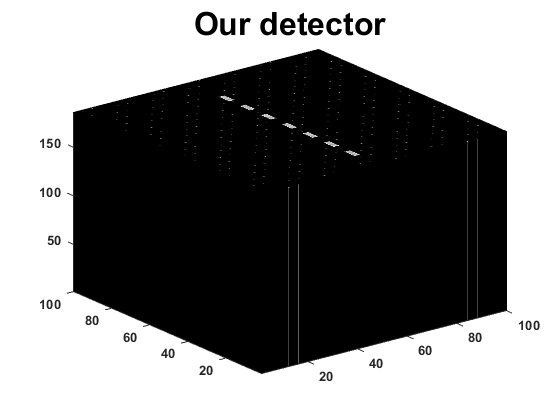}
\endminipage

 \begin{center}
\large$\bf \boldsymbol{\alpha} = 0.8$
\end{center}
\minipage{0.54\textwidth}
  \includegraphics[width=\linewidth]{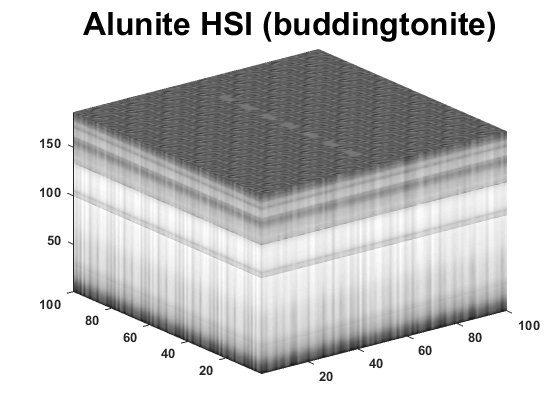}
\endminipage\hfill
\minipage{0.54\textwidth}
  \includegraphics[width=\linewidth]{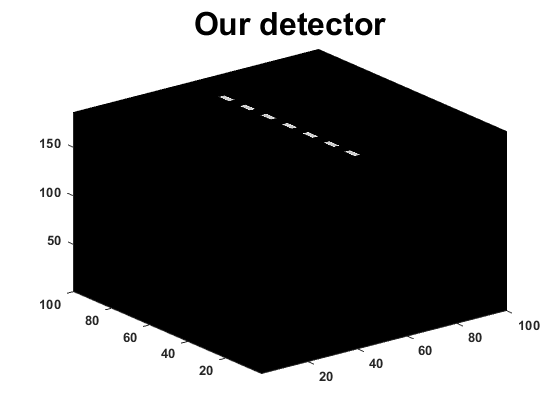}
\endminipage\hfill
\minipage{0.54\textwidth}
  \includegraphics[width=\linewidth]{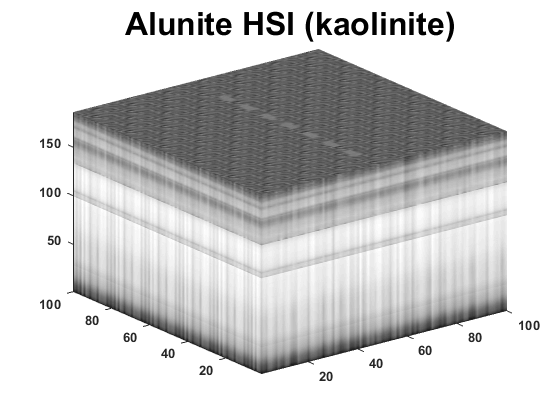}
\endminipage\hfill
\minipage{0.54\textwidth}
  \includegraphics[width=\linewidth]{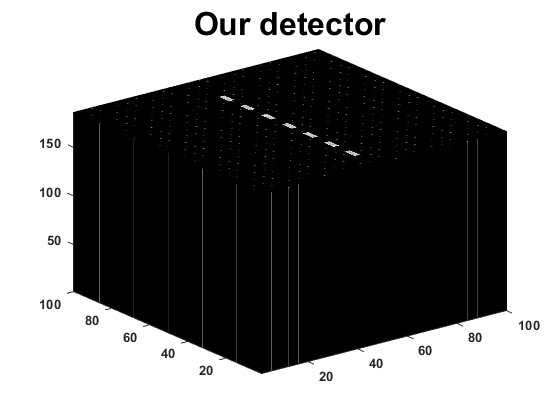}
\endminipage
\end{figure}

 \begin{figure}[!tbp]
 \begin{center}
\large$\bf \boldsymbol{\alpha} = 0.5$
\end{center}
\minipage{0.54\textwidth}
  \includegraphics[width=\linewidth]{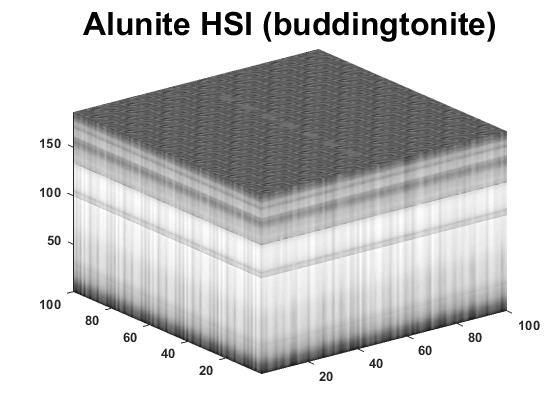}
\endminipage\hfill
\minipage{0.54\textwidth}
  \includegraphics[width=\linewidth]{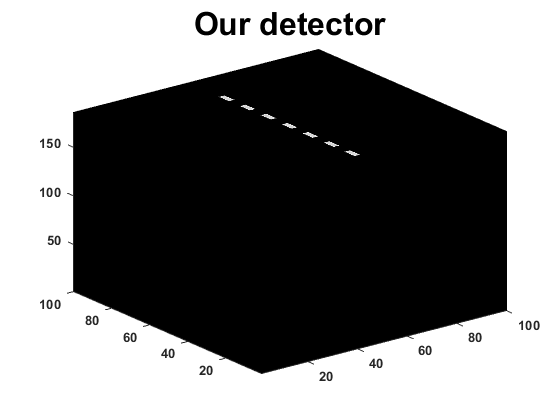}
\endminipage\hfill
\minipage{0.54\textwidth}
  \includegraphics[width=\linewidth]{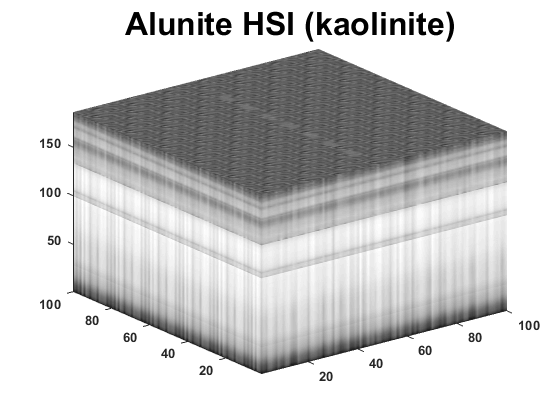}
\endminipage\hfill
\minipage{0.54\textwidth}
  \includegraphics[width=\linewidth]{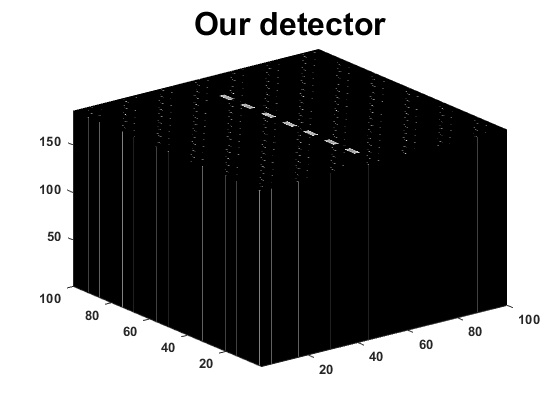}
\endminipage

 \begin{center}
\large$\bf \boldsymbol{\alpha} = 0.3$
\end{center}
\minipage{0.54\textwidth}
  \includegraphics[width=\linewidth]{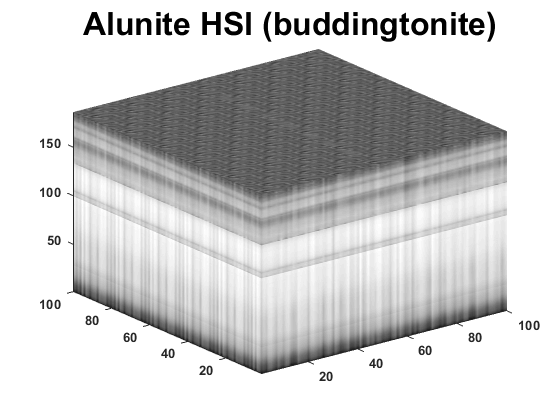}
\endminipage\hfill
\minipage{0.54\textwidth}
  \includegraphics[width=\linewidth]{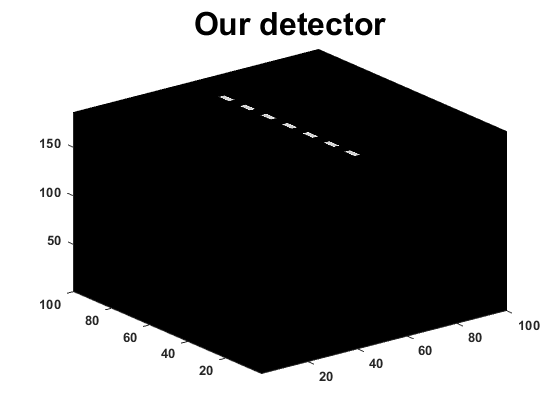}
\endminipage\hfill
\minipage{0.54\textwidth}
  \includegraphics[width=\linewidth]{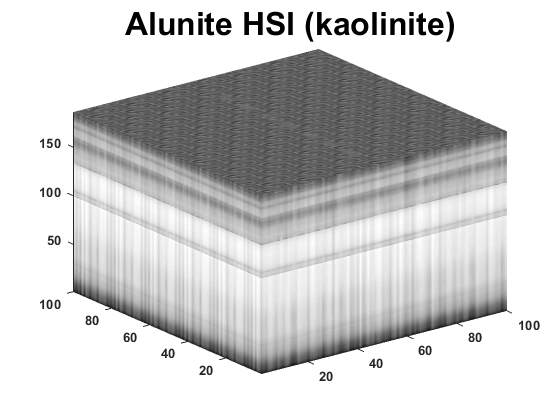}
\endminipage\hfill
\minipage{0.54\textwidth}
  \includegraphics[width=\linewidth]{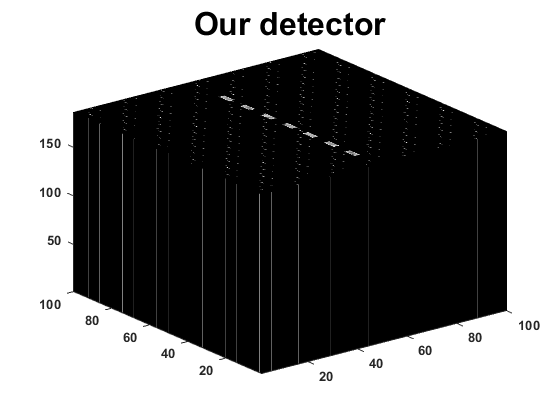}
\endminipage
\end{figure}

 \begin{figure}[!tbp]
  \begin{center}
\large$\bf \boldsymbol{\alpha} = 0.1$
\end{center}
\minipage{0.54\textwidth}
  \includegraphics[width=\linewidth]{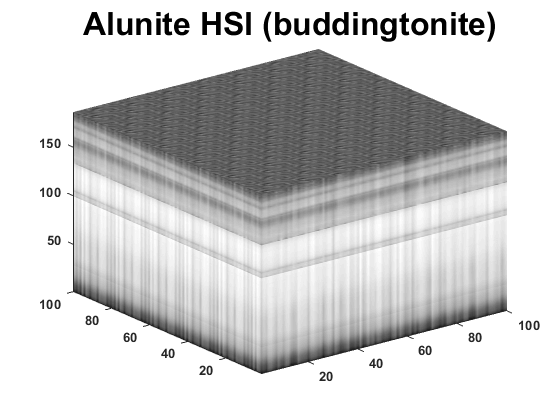}
\endminipage\hfill
\minipage{0.54\textwidth}
  \includegraphics[width=\linewidth]{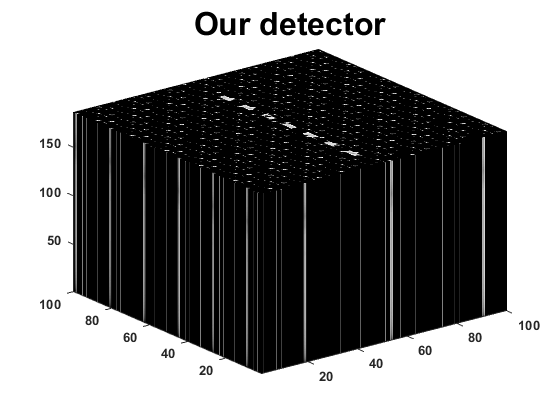}
\endminipage\hfill
\minipage{0.54\textwidth}
  \includegraphics[width=\linewidth]{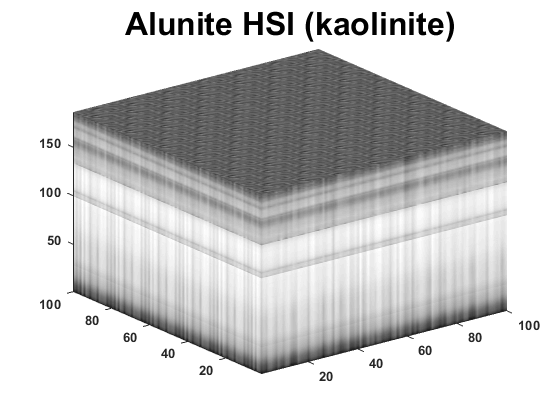}
\endminipage\hfill
\minipage{0.54\textwidth}
  \includegraphics[width=\linewidth]{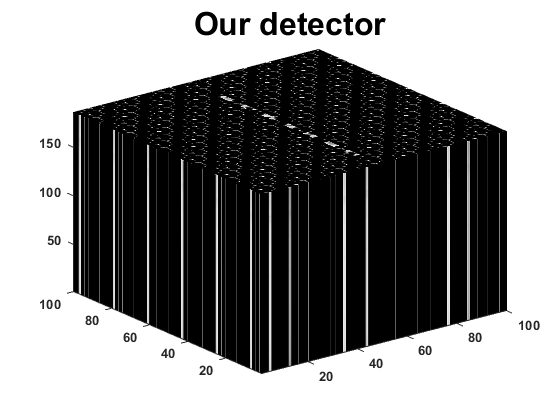}
\endminipage

 \begin{center}
\large$\bf \boldsymbol{\alpha} = 0.05$
\end{center}
\minipage{0.54\textwidth}
  \includegraphics[width=\linewidth]{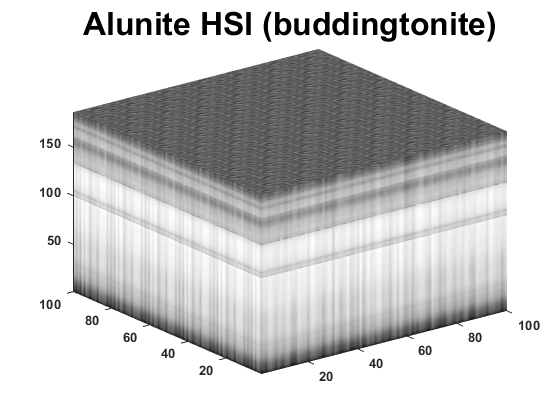}
\endminipage\hfill
\minipage{0.54\textwidth}
  \includegraphics[width=\linewidth]{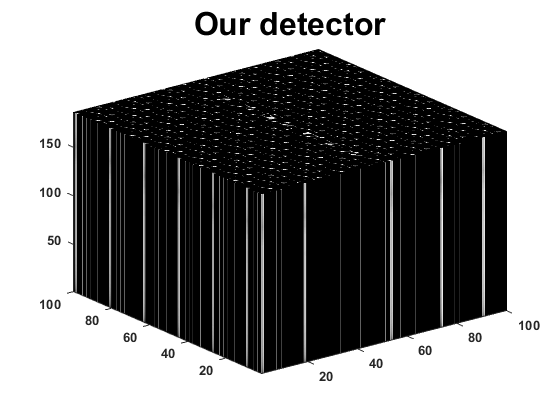}
\endminipage\hfill
\minipage{0.54\textwidth}
  \includegraphics[width=\linewidth]{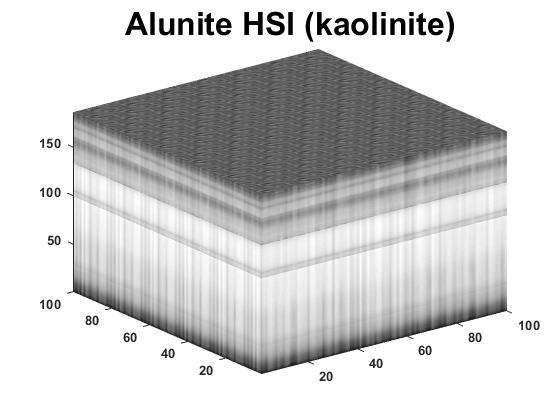}
\endminipage\hfill
\minipage{0.54\textwidth}
  \includegraphics[width=\linewidth]{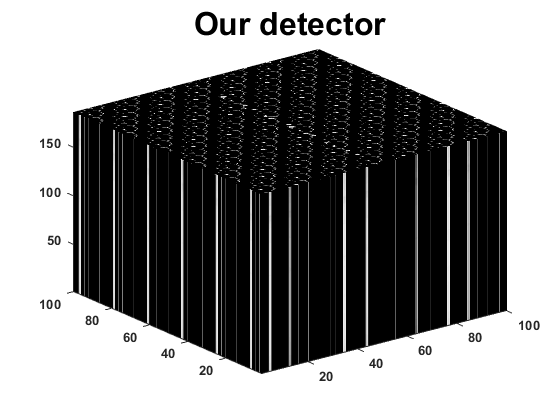}
\endminipage
\end{figure}

 \begin{figure}[!tbp]
 \begin{center}
\large$\bf \boldsymbol{\alpha} = 0.02$
\end{center}
\minipage{0.54\textwidth}
  \includegraphics[width=\linewidth]{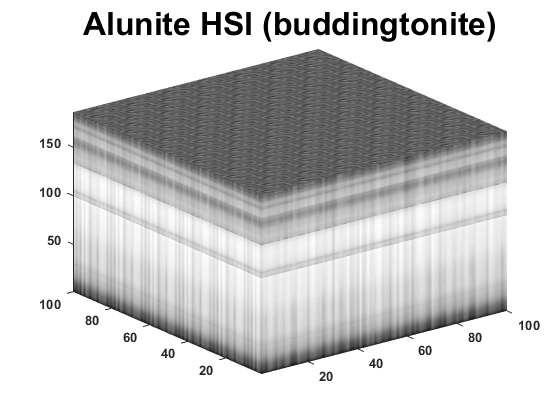}
\endminipage\hfill
\minipage{0.54\textwidth}
  \includegraphics[width=\linewidth]{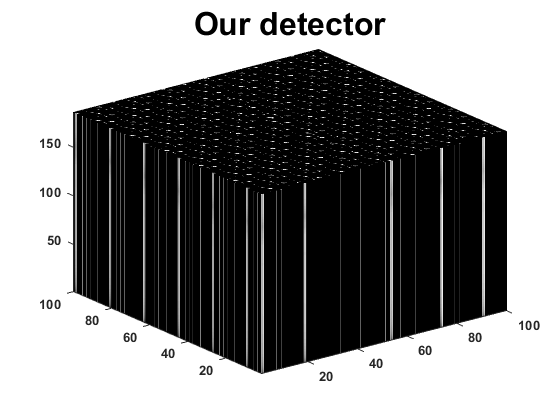}
\endminipage\hfill
\minipage{0.54\textwidth}
  \includegraphics[width=\linewidth]{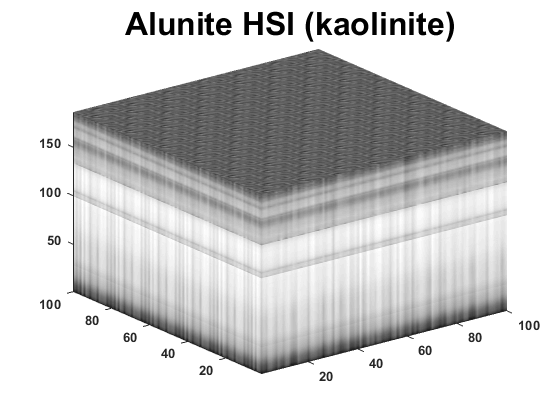}
\endminipage\hfill
\minipage{0.54\textwidth}
  \includegraphics[width=\linewidth]{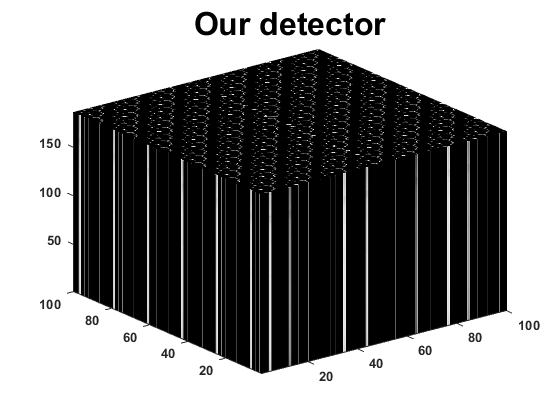}
\endminipage

 \begin{center}
\large$\bf \boldsymbol{\alpha} = 0.01$
\end{center}
\minipage{0.54\textwidth}
  \includegraphics[width=\linewidth]{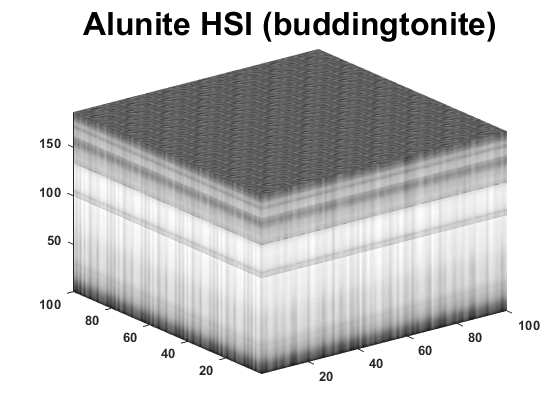}
\endminipage\hfill
\minipage{0.54\textwidth}
  \includegraphics[width=\linewidth]{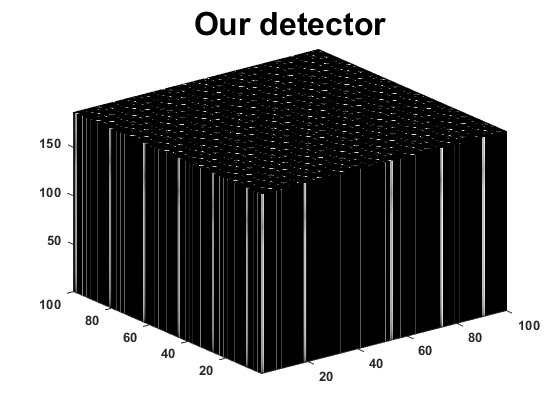}
\endminipage\hfill
\minipage{0.54\textwidth}
  \includegraphics[width=\linewidth]{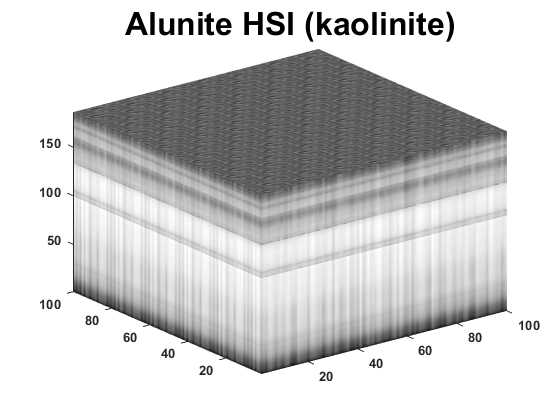}
\endminipage\hfill
\minipage{0.54\textwidth}
  \includegraphics[width=\linewidth]{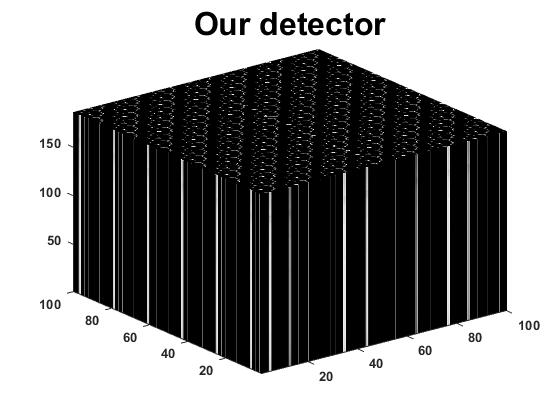}
\endminipage
\caption{Evaluation of our detector $(\mathbf{A}_t\mathbf{C})^T$ for detecting the buddingtonite and kaolinite target (for $\alpha \in [0.01, \, 1]$) when $\mathbf{A}_t$ is contsructed from target samples in the online spectral libraries.}\label{fig:detection_evaluations}
\end{figure}

\newpage
 
\section{Conclusion and Future Work}
 In this chapter, the well-known robust principal component analysis (RPCA) is exploited for target detection in hyperspectral imagery. By making assumptions similar to those used in RPCA, a given hyperspectral image (HSI) has been decomposed into the sum of a low-rank background HSI and a sparse target HSI that only contains the targets (with the background is suppressed) \cite{8677268}. In order to alleviate the inadequacy of RPCA on distinguishing the true targets from the background, we have incorporated into the RPCA imaging, the prior target information that can often be provided to the user. In this regard, we have constructed a pre-learned target dictionary $\mathbf{A}_t$, and thus, the given HSI is decomposed as the sum of a low-rank background HSI $\mathbf{L}$ and a sparse target HSI $\left(\mathbf{A}_t\mathbf{C}\right)^T$, where $\mathbf{C}$ is a sparse activation matrix.

In this work, the sparse component $\left(\mathbf{A}_t\mathbf{C}\right)^T$ was only the object of interest and used directly for the detection. More precisely, the targets are deemed to be present at the non-zero entries of the sparse target HSI.
Hence, a novel target detector is developed, which is simply a sparse HSI generated automatically from the original HSI, but containing only the targets of interest with the background is suppressed.

The detector is evaluated on real experiments, and the results of which demonstrate its effectiveness for hyperspectral target detection, especially on detecting targets that have overlapping spectral features with the background.

The $l_1$ norm regularizer, a continuous and convex surrogate, has been studied extensively in the literature \cite{Tibshirani96, 10.2307/3448465} and has been applied successfully to many applications including signal/image processing, biomedical informatics, and computer vision \cite{doi:10.1093/bioinformatics/btg308, 4483511, Beck:2009:FIS:1658360.1658364, 4799134, Ye:2012:SMB:2408736.2408739}. Although the $l_1$ norm based sparse learning formulations have achieved great success, they have been shown to be suboptimal in many cases \cite{CandAs2008, Zhanggg, zhang2013} since the $l_1$ is still too far away from the ideal $l_0$ norm. To address this issue, many non-convex regularizers, interpolated between the $l_0$ norm and the $l_1$ norm, have been proposed to better approximate
the $l_0$ norm.  They include $l_q$ norm ($0<q<1$) \cite{Fourc1624}, Smoothly Clipped Absolute Deviation \cite{Fan01}, Log-Sum Penalty \cite{Candes08}, Minimax Concave Penalty \cite{zhang2010aa}, Geman Penalty \cite{392335, 5153291}, and Capped-$l_1$ penalty \cite{Zhanggg, zhang2013, Gong12}. 
\begin{svgraybox}
In this regard, from problem \eqref{eq:convex_model}, it will be interesting to use other proxies than the $l_{2,1}$ norm, closer to $l_{2,0}$, in order to probably alleviate the $l_{2,1}$ artifact and also the manual selection problem of both $\tau$ and $\lambda$.
But although the non-convex regularizers (penalties) are appealing in sparse learning, it remains a very big challenge to solve the corresponding non-convex optimization problems.
\end{svgraybox}

\begin{acknowledgement}
The authors would greatly thank Dr.~Gregg A. Swayze from the United States Geological Survey (USGS) for his time in providing them helpful remarks about the cuprite data and especially on the buddingtonite, alunite, and kaolinite minerals. They would also like to thank the handling editors (Prof. Saurabh Prasad and Prof. Jocelyn Chanussot) and some other anonymous reviewers for the careful reading and helpful remarks/suggestions.
\end{acknowledgement}

\bibliographystyle{IEEEbib}
\bibliography{biblio_these}

\end{document}